\documentclass[12pt,letterpaper]{article}
\usepackage{amssymb}
\usepackage{amsmath}
\usepackage{amscd}
\usepackage{graphicx}
\usepackage{tikz}
\usepackage{subfigure}
\usepackage{array}
\usepackage{pdflscape}

\def\be{\begin{equation}}
\def\ee{\end{equation}}
\def\bea{\begin{eqnarray}}
\def\eea{\end{eqnarray}}
\def\bse{\begin{subequations}}
\def\ese{\end{subequations}}

\def\fracm#1#2{\hbox{\large{${\frac{{#1}}{{#2}}}$}}}

\newcommand{\BldmTH }[1]{\mbox{\boldmath$#1$}}

\def\half{\frac12}

\def\vCent#1{\vcenter{\hbox{\hss#1\hss}}}

\def\bna{\bql\begin{array}{rcl}}
\def\ena{\end{array}\eql}
\def\bnn{\beq\begin{array}{rcl}}
\def\enn{\end{array}\ee}
\def\bet{\begin{tabular}}
\def\bsf{\sffamily\bfseries}

\def\pp{{\vphantom{+}\smash{\mathchar'75\mkern-9mu|\mkern5mu}}}
\def\mm{{=\,}}
\def\4{\text{\bsf-}}

\def\hi{{\hat\imath}}

\def\hj{{\hat\jmath}}

\def\fc#1#2{\relax\ifmmode{\scriptstyle\frac{#1}{#2}} 
                    \else$\scriptstyle\frac{#1}{#2}$\fi}    

\definecolor{Red}    {rgb}{1.00,0.00,0.00} 
\definecolor{Green}  {rgb}{0.00,0.75,0.00} 
\definecolor{Blue}   {rgb}{0.00,0.00,1.00} 
\definecolor{Orange} {rgb}{1.00,0.67,0.00} 
\definecolor{Purple} {rgb}{0.50,0.00,0.50} 
\definecolor{Gold}   {rgb}{1.00,0.90,0.00} 
\definecolor{Magenta}{rgb}{1.00,0.00,1.00} 
\definecolor{Turque} {rgb}{0.00,0.90,0.90} 
\definecolor{Seaweed}{rgb}{0.00,0.25,0.00} 
\definecolor{Brown}  {rgb}{0.50,0.13,0.00} 
\definecolor{Cobalt} {rgb}{0.00,0.00,0.50} 
\definecolor{Sage}   {rgb}{0.00,0.50,0.38} 
\definecolor{grey1}  {rgb}{0.20,0.20,0.20} 
\definecolor{grey2}  {rgb}{0.40,0.40,0.40} 
\definecolor{grey3}  {rgb}{0.60,0.60,0.60} 
\definecolor{grey4}  {rgb}{0.80,0.80,0.80} 
\definecolor{grey5}  {rgb}{0.90,0.90,0.90} 
\def\C#1#2{{\ifcase#1\or
             \color{Red}\or\color{Green}\or\color{Blue}\or
              \color{Orange}\or\color{Purple}\or\color{Gold}\or
             \color{Magenta}\or\color{Turque}\or\color{Seaweed}\or
               \color{Brown}\or\color{Cobalt}\or\color{Sage}\or
                 \color{grey1}\or\color{grey2}\or\color{grey3}\or
                 \color{grey4}\else\color{grey5}\fi#2}}
\definecolor{gray}{rgb}{.7,.7,.7}
\def\XXX{\colorbox{yellow}{\color{red}\bf X\kern-4pt{\Large$\bs*$}\kern-4.125ptX}}
\def\pa{\partial}

\def\a{{\alpha}}
\def\b{{\beta}}
\def\g{{\gamma}}
\def\s{{\sigma}}

\def\[{\left[}
\def\]{\right]}

\font\ro=cmsy10                          
\def\kcr{{\hbox{\ro \char'170}}}                
\def\ktl{{\hbox{\ro \char'170}}}        
\def\ktr{{\hbox{\ro \char'170}}}        
\def\kbl{{\hbox{\ro \char'170}}}        
\def\kbr{{\hbox{\ro \char'170}}}        

\def\rI{{{}_{\rm I}}}
\def\rJ{{{}_{\rm J}}}

\def\hj{{\hat\jmath}}
\def\hk{{\hat k}}
\def\hi{{\hat\imath}}

\parskip=\medskipamount                 
\thicklines                         

\newskip\humongous \humongous=0pt plus 1000pt minus 1000pt
\def\caja{\mathsurround=0pt}
\def\eqalign#1{\,\vcenter{\openup2\jot \caja
        \ialign{\strut \hfil$\displaystyle{##}$&$
        \displaystyle{{}##}$\hfil\crcr#1\crcr}}\,}
\newif\ifdtup

\def\border{                                            
        \setlength{\unitlength}{1mm}
        \newcount\xco
        \newcount\yco
        \xco=-21
        \yco=12
        \begin{picture}(140,0)
        \put(\xco,\yco){$\ktl$}
        \advance\yco by-1
        {\loop
        \put(\xco,\yco){$\kcr$}
        \advance\yco by-2
        \ifnum\yco>-240
        \repeat
        \put(\xco,\yco){$\kbl$}}
        \xco=158
        \yco=12
        \put(\xco,\yco){$\ktr$}
        \advance\yco by-1
        {\loop
        \put(\xco,\yco){$\kcr$}
        \advance\yco by-2
        \ifnum\yco>-240
        \repeat
        \put(\xco,\yco){$\kbr$}}
        \put(-19.75,13){\tiny **University of Maryland * Center for String and
         Particle  Theory * Physics Department**University of Maryland * Center
        for String and Particle  Theory** }
        \put(-19.75,-241.5){\tiny **University of Maryland * Center for String and
         Particle  Theory * Physics Department**University of Maryland * Center
        for String and Particle  Theory** }
        \end{picture}
        \par\vskip-8mm}

\def\headpic{                                           
        \indent
        \setlength{\unitlength}{.4mm}
        \thinlines
        \par
        \begin{picture}(29,16)
        \put(165,16){\line(1,0){4}}
        \put(170,16){\line(1,0){4}}
        \put(180,16){\line(1,0){4}}
        \put(175,0){\line(1,0){4}}
        \put(180,0){\line(1,0){4}}
        \put(185,0){\line(1,0){4}}
        \put(169,0){\line(0,1){16}}
        \put(170,0){\line(0,1){16}}
        \put(179,0){\line(0,1){16}}
        \put(180,0){\line(0,1){16}}
        \put(184,0){\line(0,1){16}}
        \put(185,0){\line(0,1){16}}
        \put(169,16){\oval(8,32)[bl]}
        \put(170,16){\oval(8,32)[br]}
        \put(179,0){\oval(8,32)[tl]}
        \put(185,0){\oval(8,32)[tr]}
        \end{picture}
        \par\vskip-6.5mm
        \thicklines}
\def\endtitle{\end{quotation}\newpage}                  

\begin{document}

\border\headpic {\hbox to\hsize{\today \hfill
{PP-012-016}}}
\par \noindent
{ \hfill {}}
\par

\par

\setlength{\oddsidemargin}{0.5in}
\setlength{\evensidemargin}{-0.5in}
\begin{center}
\vglue .10in
{\large\bf A Computer Algorithm For \\ \vskip0.1in
Engineering Off-Shell Multiplets
\\ \vskip0.2in
With Four Supercharges On The World Sheet}
\\[.5in]
Keith\, Burghardt\footnote{keith@umd.edu}, and
S.\, James Gates, Jr.\footnote{gatess@wam.umd.edu}
\\[1.2in]

{\it Center for String and Particle Theory\\
Department of Physics, University of Maryland\\
College Park, MD 20742-4111 USA}\\[1.4in]

{\bf ABSTRACT
}
\\[.005in]
\end{center}
\begin{quotation}
{We present an adinkra-based computer algorithm 
implemented in a Mathematica code and use it in a limited
demonstration of how to engineer off-shell multiplets 
with four supercharges on the world sheet. Using one of the 
outputs from this algorithm, we present evidence for the 
unexpected discovery of a previously unknown 8 - 8 
representation of $\cal N$ = 2  world sheet supersymmetry.  
As well, we uncover a menagerie of ($p$, $q$) = (3, 1) world 
sheet supermultiplets.
} \endtitle

\section{Introduction}
$~~~$

Historically, there have been two approaches used for the discovery
of off-shell supersymmetrical multiplets and these are:  \newline \indent
(a.) the component method, and \newline \indent
(b.) the superfield method.  \newline \noindent
The former dates to the earliest realizations of supersymmetrical 
representations \cite{CFmeth}.  It relies on intuition to make well-motivated
guesses as to the field content of a supermultiplet and these are
verified by checking for the closure of a set of SUSY variations.
The latter dates to introduction of superfields \cite{SFmeth}. In
this method, one relies on intuition to make well-motivated
guesses as the structure of a set of super-differential equations
to define an irreducible off-shell multiplet.

{\em {The use of intuition and guesses in both of these methods is indicative
of a need for a comprehensive theory of off-shell linear representations
of superymmetry.}}

It is the purpose of this work to introduce a new and third method for
the discovery of off-shell linear supersymmetrical representations.  In the 
new approach, intuition and guesses are replaced by a computer-enabled
search of the 1D zoo of off-shell linear supersymmetrical representations
provided by adinkras \cite{Adinkra}.  In the following, we will describe an approach
that relies on recent advances in understanding the relation of adinkras
to supersymmetrical field theories on the world sheet in order to created an algorithm.
It is capable of systematically scanning the adinkra zoo for representations that
can be lifted to 2D theories.  Initially, this was undertaken to
simply validate this approach in a small domain of adinkras.  Quite 
unexpectedly, we seem to have found a new representation.  If this
interpretation stands up to additional investigation, it will mark the
first time a new off-shell linear representation has been discovered
by a systemic deductive means..

Adinkras have been shown to completely describe 
off-shell one dimensional linear realizations of supersymmetry in graphical 
form. However, we have long argued \cite{ENUF} that their applicability 
extends to $\cal N$-extended supersymmetrical representations in all 
dimensions.  The basis for this suggestion is the proposal that adinkras 
are to off-shell supersymmetric representation theory as SU(3) weight 
space diagrams are to hadronic representations.  It is well known that 
SU(3) weight space diagrams do not just apply to the fundamental 
representations, but to all representations.
 
It was predicted in the work of Ref.\  \cite{ENUF} that there must exist a property 
of ``SUSY holography" by which higher dimensional off-shell SUSY
representations are holographically stored in the representation theory
of one dimensional SUSY quantum mechanical systems.  In other words,
adinkras must be holograms for higher dimensional SUSY representations.

The first actual demonstration of this property was shown in the work of  Ref.\  \cite{DimUp},
where a logically consistent mathematical manner for doing so was presented 
in some examples.  Furthermore, a set of requirements for dimensional 
extensions more generally were given.  However, this approach is calculationally 
demanding as it basically requires checks of compatibility between equations.   

Quite recently and using entirely different methods \cite{Plato}, it was shown 
how 1D $\cal N$ = 4 valise adinkras holographically encode the data required
to reconstruct the (1, 3) Clifford bundle required for four dimensional fermions
in some 4D,  $\cal N$ = 1 supermultiplets.  As valise adinkras
form the simplest of one dimensional supersymmetric representations,
this observation diminishes some of the effort surrounding node-raising 
analysis in the work of  Ref.\  \cite{DimUp}.  This will be discussed in 
a specific result at the end of chapter six.

The search for computationally efficient methods for the derivation, from 
adinkra-based descriptions, of SUSY multiplets in greater than one dimension
continued.  This resulted in Ref.\  \cite{Bowtie}, which found that 
a special class of paths (known as ``two-color ambidextrous bow ties" -  see Fig.\ \# 1) 
within adinkras act as obstructions to their extension from one dimensional 
to two dimensional systems.  

We believe the work of Ref.\  \cite{Bowtie} will stand any challenge to the
idea that SUSY holography exists, at least in the case of going from one
to two bosonic dimensions.  As off-shell two dimensional SUSY representations 
are contained within four dimensional SUSY representations, work showing 
dimensional extension from one to two dimensions also supports the contention 
about SUSY holography more generally.  

{\em {With the discovery of the  ``no two-color ambidextrous bow tie"
theorem there now exists a proof that adinkras have been shown to be holograms
for off-shell two dimensional  linear realizations of supersymmetry.}}

The absence of these special ``two-color ambidextrous bow tie" paths acts 
as a filter.  A problem has remained, however, on how to efficiently determine 
whether adinkras are liftable without having to visually inspect 2-color paths 
in adinkras looking for the obstructive paths.  This is due to the shear number 
of paths through which one needs to search.  

In this work, we present an efficient algorithm which can determine liftability for 
arbitrary adinkras of arbitrary size, using two groups of matrices that we shall 
introduce and call by the name ``color matrices."  These will be denoted by the symbols
$\mathcal{B}{}_{{}_{\rm I}L}$ and $\mathcal{B}{}_{{}_{\rm I}R}$.  These matrices 
are then used to construct ``color block matrices," $\mathbf{C}_{{}_{\rm I}}$, which 
describe the linear transformations on the entire vector space of bosons and 
fermions, $\mathbf{\Phi}\oplus\mathbf{\Psi}$ under the action of the ${\rm 
D}{}_{{}_{\rm I}}$ super-differential operator.

We will show that multiplying two color matrices together, $\mathcal{B}
{}_{{}_{\rm I}L} \mathcal{B}{}_{{}_{\rm J}R}$, can be used to count the number of 
bow ties for any two-color combination, $(I,J)$. Knowing whether bow ties exist 
for specific color combinations is all one needs to determine if adinkras have 
ambidextrous bow ties and are therefore non-liftable. In addition, we present 
results concerning the dimensional extension of 1D adinkras to 2D adinkras 
based on the tesseract adinkra and code-folded tesseract adinkra.  These 
results will demonstrate the utility of this algorithm.  This gives an explicit path by which 
one can start solely with adinkras and no other information to engineer the 
equations for off-shell supermultiplets on the world sheet.   

The paper is organized as follows: in Section 2, we describe the different types of 
liftable and non-liftable two-color paths.  In Section 3, we review the Bow Tie Theorem 
in Ref.\  \cite{Bowtie}, and bow tie graphs, which we use to determine liftability of the 
adinkra. In Section 4, we introduce the liftability algorithm, and the color matrices, 
which are essential to determine liftability.  In Section 5, we give a discussion
on why adding the notion of helicity to one dimensional adinkras triggers the
need to introduce a second bosonic coordinate and, perhaps even more interesting,
a Lorentzian structure for a world sheet.   In Section 6, we describe the results
of a search that demonstrate the algorithm's usefulness by calculating which 
ambidextrous adinkras based on the tesseract adinkra and code-folded tesseract 
are liftable,  This includes an enumeration.  Our conclusions are given in 
Section 7. We include four appendices.  In the first and second there are given
the 2D SUSY equations (equivalent to SUSY variations) for two classes of liftable 
adinkras that correspond to the $\cal {GR}$(2,2) and $\cal {GR}$(4,4)
algebras. The third appendix presents an evidence for one such adinkra
is a previously unknown representation.  The final appendix presents
the Mathematica code used for the search for liftable adinkras. 

\section{On Bow Tie \& Diamond Adinkras}

 $~~~$ No matter how many colors appear in an arbitrarily complicated adinkra, 
 the work of Ref.\  \cite{Bowtie} implies that only two-color closed paths are relevant
 to the dimensional extension of supersymmetrical representations from those
 that depend on a single bosonic coordinate to those that depend on two bosonic 
 coordinates.  There are only two types of such paths as shown in Fig.\ \# 1.
\begin{figure}[!htbp]
   \centering
       \subfigure[]{\label{f:bowtie}
    \includegraphics[width=0.4\columnwidth]{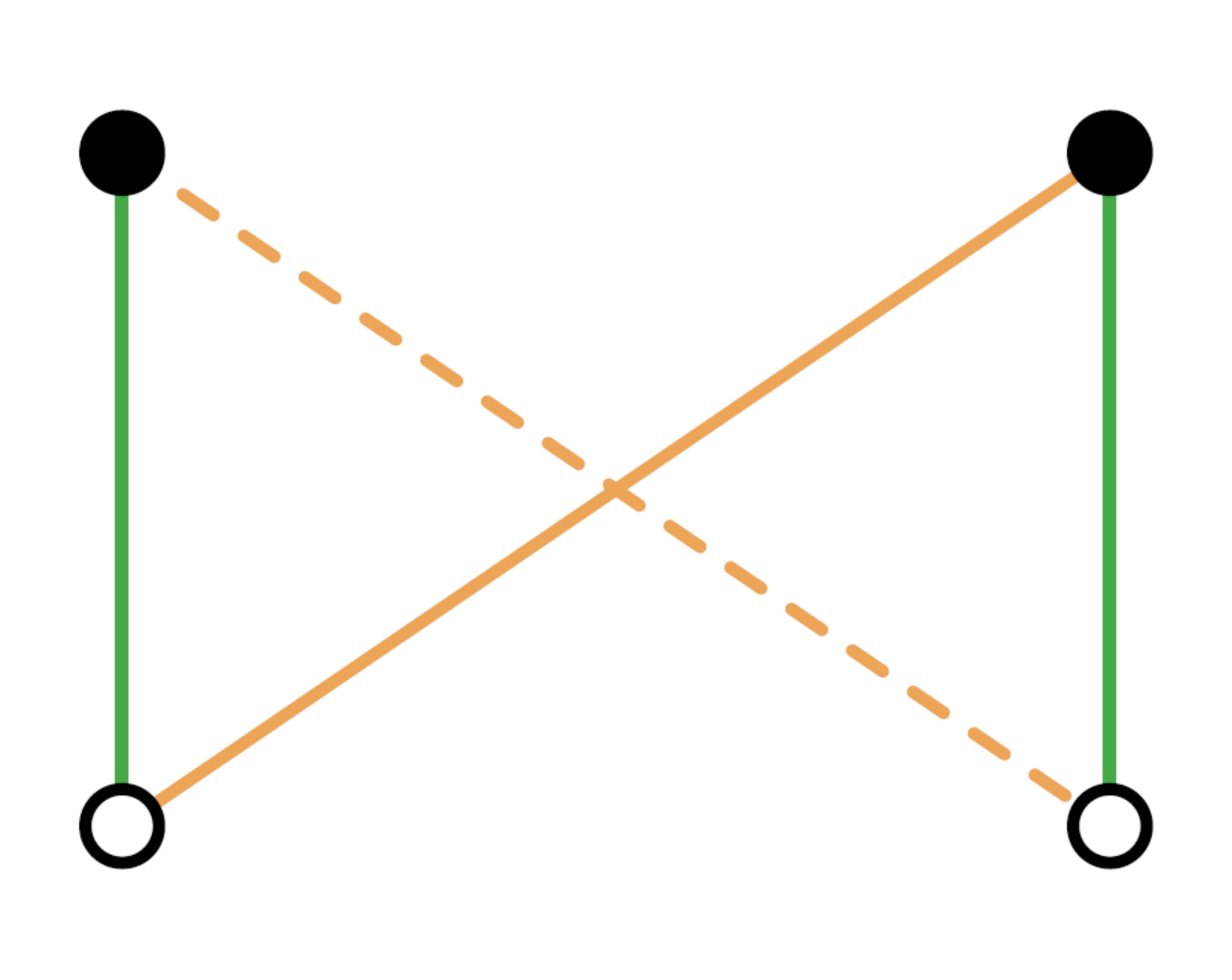}} 
        \quad
    \subfigure[]{\label{f:diamond}
    \includegraphics[width=0.4\columnwidth]{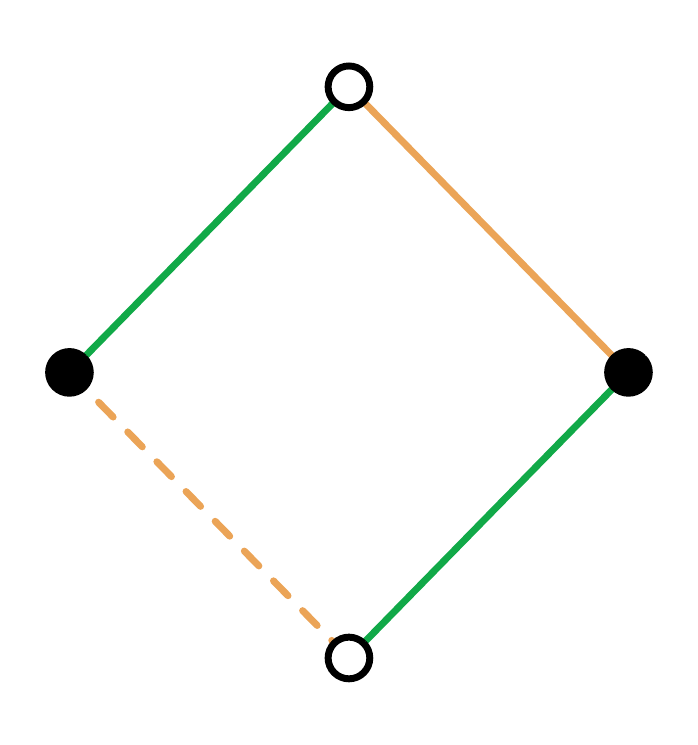}}
    \quad
    \label{f:diambt}
    \caption{The bow tie two-color path (left) and the diamond
     two-color path (right).}
\end{figure}
The main difference between one dimensional spinors and two dimensional
spinors arises from the fact that only the latter possess a helicity, i.\ e.\ a two
dimensional spinor can be `spin-up' ($+\half$) or `spin-down' ($-\half$). 
Operationally, this means that if one begins with a one dimensional spinor
derivative operator, ${\rm D}{}_{{}_{\rm I}}$, it may be regarded as having arisen 
from the projection of either ${\rm D}{}_{+{}_{\rm I}}$ or ${\rm D}{}_{-{}_{\rm I}}$. 
So as a minimal requirement for extending from one dimensions to two dimensions, 
an operator Spin(${\rm D}{}_{{}_{\rm I}}$) must be introduced.  This operator assigns 
a value of the two-dimensional helicity (either ${\rm D}{}_{+{}_{\rm I}}$ or ${\rm 
D}{}_{-{}_{\rm I}}$) to the one-dimensional spinor ${\rm D}{}_{{}_{\rm I}}$.

Ambidextrous two-color bowtie adinkras are bowtie adinkras where the assignment
of spin to the two distinct colors is such that one has spin up and the other has spin 
down. Unidextrous two-color bowtie adinkras are adinkras where the assignment 
of spin to the two distinct colors is such that both have either spin up or both have
spin down.   Ambidextrous and unidextrous bow ties look the same from the point 
of view of spinors in a space with a single bosonic dimension.  In a space of two 
bosonic dimensions things change due to the operator Spin(${\rm D}{}_{{}_{\rm I}}$).  

For example, the diamond adinkra in  Fig.\ \# 1 under the action of all possible 
possible choices of the spin gives rise to the four graphs that we show in Fig.\ \# 2. 
As the links in the first two of these adinkras are associated with both of the two 
possible spin states, they are ambidextrous.  In the third and fourth adinkras.
the links are associated solely with spin up (third adinkra) or solely with spin
down (fourth adinkra).  Hence, they are unidextrous diamond adinkras.
 $$
\vCent
 {\setlength{\unitlength}{1mm}
  \begin{picture}(-20,0)
   \put(-74,-28){\includegraphics[width=1.2in]{figures/Diamond}}
   \put(-69,-4){$+$} \put(-51,-4){$-$}      
  \put(-69,-22){$-$} \put(-51,-22){$+$}  
  \end{picture}}
$$ 
 $$
\vCent
 {\setlength{\unitlength}{1mm}
  \begin{picture}(-20,0)
   \put(-34,-18.5){\includegraphics[width=1.2in]{figures/Diamond}}
   \put(-28,5,5){$-$} \put(-12,5.5){$+$}      
  \put(-28,-12.5){$+$} \put(-12,-12.5){$-$}  
  \end{picture}}
$$
 $$
\vCent
 {\setlength{\unitlength}{1mm}
  \begin{picture}(-20,0)
   \put(6,-10.5){\includegraphics[width=1.2in]{figures/Diamond}}
   \put(11,14){$+$} \put(29,14){$+$}      
  \put(11,-4){$+$} \put(29,-4){$+$}  
  \end{picture}}
$$ 
 $$
\vCent
 {\setlength{\unitlength}{1mm}
  \begin{picture}(-20,0)
   \put(43,-2.5){\includegraphics[width=1.2in]{figures/Diamond}}
   \put(48,22){$-$} \put(66,22){$-$}    
   \put(48,4){$-$} \put(66,4){$-$}  
   \put(-56,-10){Figure 2: Ambidextrous and unidextrous diamond adinkras.}  
  \end{picture}}
$$
\vskip0.4in
 \noindent 
In a similar, manner the bow tie adinkra in  Fig.\ \# 1 under the action 
of all possible choices of the spin gives rise to four graphs below.
 $$
\vCent
 {\setlength{\unitlength}{1mm}
  \begin{picture}(-20,0)
   \put(-74,-25){\includegraphics[width=1.2in]{figures/BowtieA}}
   \put(-65,-6){$-$} \put(-55,-6){$-$}      
  \put(-76,-13){$+$} \put(-45,-13){$+$}  
  \end{picture}}
$$ 
 $$
\vCent
 {\setlength{\unitlength}{1mm}
  \begin{picture}(-20,0)
   \put(-34,-15){\includegraphics[width=1.2in]{figures/BowtieA}}
   \put(-25,4){$+$} \put(-15,4){$+$}      
  \put(-36,-3){$-$} \put(-4,-3){$-$}  
  \end{picture}}
$$
 $$
\vCent
 {\setlength{\unitlength}{1mm}
  \begin{picture}(-20,0)
   \put(6,-6.5){\includegraphics[width=1.2in]{figures/BowtieA}}
   \put(15,12){$+$} \put(24,12){$+$}      
  \put(4,5.25){$+$} \put(35,5){$+$}  
  \end{picture}}
$$ 
 $$
\vCent
 {\setlength{\unitlength}{1mm}
  \begin{picture}(-20,0)
   \put(43,1.5){\includegraphics[width=1.2in]{figures/BowtieA}}
   \put(51,20){$-$} \put(62,20){$-$}    
      \put(41,13.5){$-$} \put(72,13.5){$-$}    
  \end{picture}}
$$
\centerline{{Figure 3: Ambidextrous (left) and unidextrous (right) bow tie adinkras.}}
\noindent
It is the {\em a} {\em {priori}} appearance of the ambidextrous two-color bow tie 
adinkras (as in the two leftmost graphs above) that prevent such adinkras from 
describing supersymmetrical systems whose nodes depend on two bosonic 
coordinates.  In adinkras of more colors, it is this subset of graphs that prevents
such adinkras from describing supersymmetrical systems whose nodes are
permitted to depend on two bosonic coordinates.

\section{Liftability and Bow Ties}

$~~~$ The Bow Tie Theorem \cite{Bowtie} is based on the possibility of defining
a graph theoretic quantity $\mathcal{B}_{N}$ which is used to explicitly 
determine `liftability' of two-color paths.  The formal mathematical statement of the 
discussion in this chapter can all be found in the work of Ref.\  \cite{Bowtie}.  Here we 
will take a different track and approach this from arguments familiar to physicists. 

Let us begin using physical analogies to understand the ``Bow Tie Theorem.''  
 We can imagine the images in Fig.\ \# 1 represent electrical circuits.  To answer 
 the question of the voltage difference measured across any node, it would 
 suffice to use Kirchoff's Voltage Law.  As an integral this takes the form
\be
\eqalign{
{\cal V} ~=~ - \,
\oint  \, {\vec {\cal E}} \, \cdot \, d {\vec {\ell}}  ~~~,
} 
\label{KLaw}
\ee
where the closed integral begins on one side of an infinitesimal node, is taken through 
the adjoining link, follows a path that ends up on the opposite of the side of the
same node after traversing the circuit.

Next we make the replacements given by
\be
\eqalign{
{\cal V} &~~ \longrightarrow ~ {\cal B}_N  ~~~,
\cr
\oint  \, {\vec {\cal E}} \, \cdot \, d {\vec {\ell}}  &~~ \longrightarrow ~ - \, 
\sum_{links} \, {\rm Spin}({\rm D}{}_{{}_{\rm I}})
 \, (~ h_f ~-~ h_i ~)    ~~~.
} 
\label{Replacemnts}
\ee
The first of these replaces the voltage by a quantity we may call the
``Bow Tie Number'' denoted by ${\cal B}_N$.  In the second replacement,
the integral is replaced by a sum over the links in a closed path.  The
paths that are relevant are two-color minimal length circuits in the adinkra
and the links are chosen to correspond to this restriction.  The differential 
line elements are replaced by the differences in heights $ (h_f ~-~ h_i) $
of the end points of the links along a path, and the electrical field is replaced 
by a function ${\rm Spin}({\rm D}{}_{{}_{\rm I}})$ which assigns values $\pm 
\fracm 12$ to the D-operator associated with each link.  

It is well known that given a capacitance C${}_0$, there is an associated
energy
\be
\eqalign{
{\cal U} ~&=~ \fracm 12 \, {\rm C}_0 \,
\big[ \,  \oint  \, {\vec {\cal E}} \, \cdot \, d {\vec {\ell}} \, ~ \big]^2    ~~~,
} 
\label{CapaciEnergy}
\ee
but, we can use a second analogy to understand the significance of ${\cal 
B}_N$.  Let us imagine a physical system where there is an interaction 
energy between a spin field ${\vec {\cal S}} $ and differential lattice 
vectors $d {\vec {\ell}}$ such that the energy takes the form
\be
\eqalign{
{\cal U} ~&=~ \fracm 12 \, {\rm L}_0 \,
\big[ \,  \oint  \, {\vec {\cal S}} \, \cdot \, d {\vec {\ell}} \, ~ \big]^2    ~~~.
} 
\label{SpinEnergy}
\ee
A discretization of the integral immediately above, leads to an expression 
that is virtually the same as appears in (\ref{Replacemnts}) where the function
${\rm Spin}({\rm D}{}_{{}_{\rm I}})$ corresponds to the projection of the spin 
onto the differential lattice vectors.  The quantity ${\rm L}_0$ is introduced in 
order for the equation in (\ref{SpinEnergy}) to possess the correct engineering 
dimensions.

For the two-color closed paths in Fig.\ \# 1, the quantity ${\cal B}_N$ 
can be explicitly expressed as
\be
\vCent
 {\setlength{\unitlength}{1mm}
  \begin{picture}(-20,0)
   \put(-70,-4.5){\includegraphics[width=5.4in]{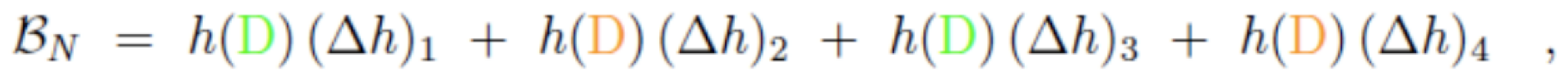}}
  \end{picture}}
\label{BTN}
\ee 
and the colorful pre-factor symbols correspond to values of $\pm \half$
depending on the choice of either spin-up or spin-down for the link of that 
color shown in Fig.\ \# 1.  In order to complete the calculation of ${\cal 
B}_N$, the orientations of the spins must be specified.  

The spin of the different colored links can either remain aligned (analogous 
to the ferromagnetic alignments of spins) or alternate between each adjacent 
links (analogous to the anti-ferromagnetic spin alignments). 
In the ``ferromagnetic case'' ${\cal B}_N$ takes the form
\be
\eqalign{
{\cal B}_N 
~&=~  \pm \, \fracm 12 \, {\big \{}  \, (\Delta h)_1   
~+~   (\Delta h)_2   ~+~   (\Delta h)_3   
~+~  (\Delta h)_4    \,  {\big \}}  ~~~,
}\label{Ferro}
\ee
and in the ``anti-ferromagnetic case'' ${\cal B}_N$ takes the form
\be
\eqalign{
{\cal B}_N 
~&=~  \pm \, \fracm 12 \, {\big \{}  \, (\Delta h)_1   
~-~   (\Delta h)_2   ~+~   (\Delta h)_3   
~-~  (\Delta h)_4    \,  {\big \}}   ~~~.
}\label{Anti-Ferro}
\ee
The third and fourth adinkras in Fig.\ \# 1 and the third and fourth adinkras 
in Fig.\ \# 2 correspond to the cases of ``ferromagnetic" conditions.  The 
first and second adinkras in Fig.\ \# 1 and the first and second adinkras 
in Fig.\ \# 2 correspond to the cases of ``anti-ferromagnetic" conditions.  

Under the ``ferromagnetic" conditions, ${\cal B}_N$ is zero for both bow tie 
and diamond adinkras.  However, under the ``anti-ferromagnetic" conditions, 
${\cal B}_N$ is zero for the diamond adinkra, but $\pm$1 for the bow tie 
adinkra\footnote{As the differences in heights in an adinkra correspond 
to differences in the engineering \newline
$~~~~~~$ dimensions of the fields related by a SUSY transformation,
these changes in the values   \newline
$~~~~~~$ 
of $h$ across a link correspond to $\pm$ 1/2 and this leads to the stated 
values for ${\cal B}_N$.}.  
 
If we think of the adinkras in Fig.\ \# 1 as the micro-states of an ensemble, 
then in the ferromagnetic regime both have the same energy as defined by	
(\ref{SpinEnergy}).  In the anti-ferromagnetic regime, only the diamond adinkra 
has the lowest energy.  So the `ground state' of the ensemble consists of
all adinkras in Fig.\ \# 2 and the unidextrous bow tie adinkras of figure
Fig.\ \# 3.  No ambidextrous two-color bow tie adinkras exist in the ground
state.

Whether a given one
dimensional adinkra can be lifted to describe a two dimensional off-shell 
multiplet is found using the Bow Tie Theorem, which states that an 
adinkra of arbitrary complexity, possessing two-color bow tie paths, cannot 
be lifted to higher dimensions if it has non-vanishing Bow Tie Number.

Clearly, if all line colors were unidextrous, the adinkra would always be liftable. 
Determining whether ``ambidextrous" adinkras are liftable is less trivial. For the 
rest of the paper, it is assumed we are lifting non-unidextrous adinkras.  

For a given adinkra, instead of checking bow tie numbers directly, it is much 
\vskip0.1in
\begin{figure}[!htbp]
   \centering
\label{f:AdnkExTess}
\includegraphics[width=0.45\columnwidth]{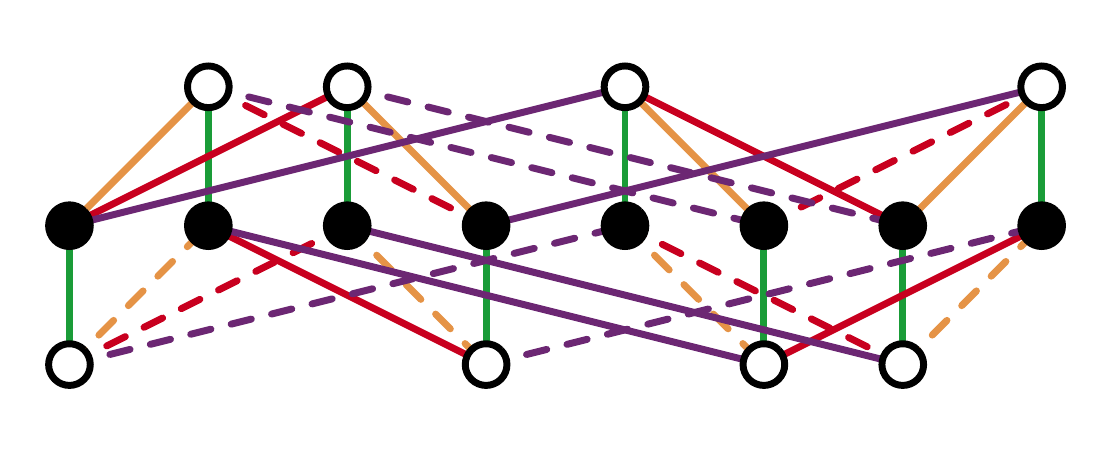}
     \end{figure} 
\centerline{{Figure 4: A liftable tesseract adinkra, with green lines that are bow 
tie free.}}
\noindent
simpler to determine liftability by checking which colors form bow ties. As long 
as there are at least 2 sets of line colors that can have opposite spin values, 
while simultaneously remaining ambidextrous bow tie-free, the adinkra is liftable.

For example, in Fig.\ \# 5, we see the valise adinkra can't be liftable because all 
the lines have bow ties. This can be symbolically shown in the ``bow tie graph" 
in Fig.\ \# 5(b), where each node represents a line color, while every line represents 
two-color bow ties between those two colors. As the graph shows, every color 
has bow ties with every other, therefore it is impossible not to create ambidextrous 
bow ties if the adinkra is non-unidextrous. For a general adinkra to be liftable, 
there must be at least two disconnected subgraphs, each of which separately 
have nodes with the same spin value to remove all ambidextrous bow ties.
$$
\vCent
{\setlength{\unitlength}{1mm}
  \begin{picture}(-20,-140)
 \put(-68,-26){\includegraphics[width=2.6in]{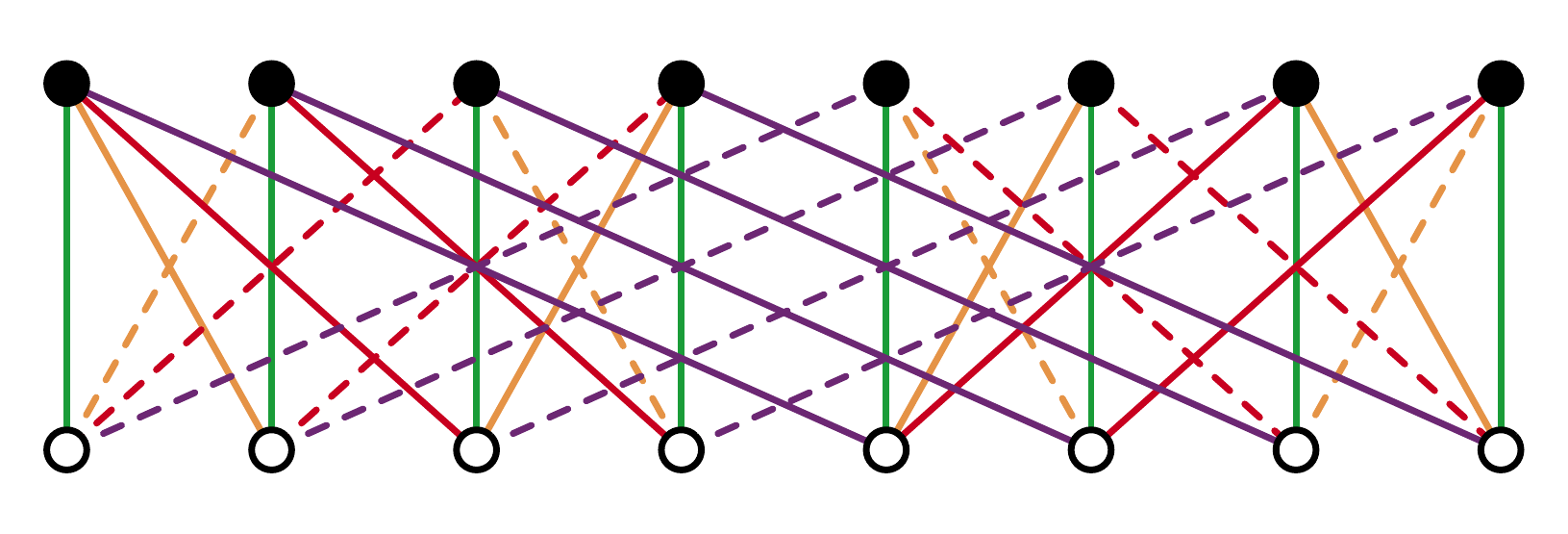}}
  \put(-36,-34){(a)}
 \put(-52,-43){Figure 5: The unliftable ambidextrous valise adinkra (left) and}
  \put(-22,-49){bow tie graph (right).}
 \end{picture}}
 \vCent
{\setlength{\unitlength}{1mm}
  \begin{picture}(-20,-140)
 \put(10,-26){\includegraphics[width=1.8in]{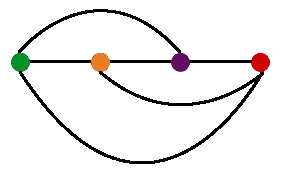}}
   \put(32,-34){(b)}
 \end{picture}}
 \nonumber
$$
\vskip1.8in  

A contrasting example appears in Fig. \# 6 where we see an example of a 
liftable 
$$
\vCent
{\setlength{\unitlength}{1mm}
  \begin{picture}(-20,-140)
 \put(-68,-20){\includegraphics[width=2.6in]{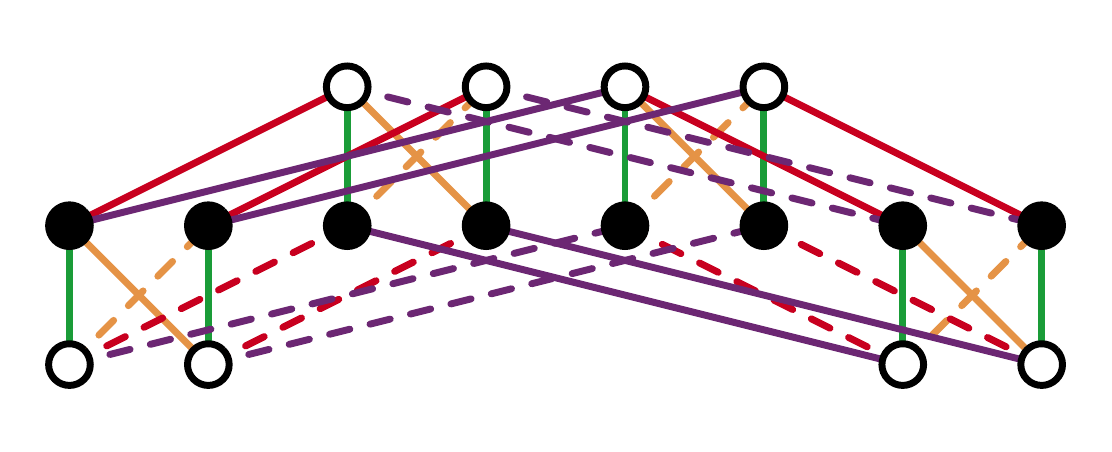}}
  \put(-36,-28){(a)}
 \put(-52,-37){Figure 6: A liftable ambidextrous valise adinkra (left) and}
  \put(-22,-43){bow tie graph (right).}
 \end{picture}}
 \vCent
{\setlength{\unitlength}{1mm}
  \begin{picture}(-20,-140)
 \put(10,-20){\includegraphics[width=1.8in]{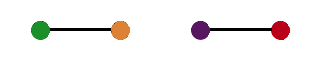}}
   \put(32,-28){(b)}
 \end{picture}}
 \nonumber
$$ \vskip1.6in \noindent
ambidextrous adinkra because the bow ties formed are green-yellow, 
as well
as red-purple. Assigning ``$-$"  to the green and yellow lines, while 
assigning ``$+$" to the red and purple lines allows the ambidextrous adinkra 
to remain liftable. This implies that the bow tie graph in Fig. \# 6(b) has two 
disconnected sub-graphs.

If an adinkra is folded, it is unnecessary to look at every line color. Instead one 
only needs to inspect the ``hypercubic" sub-adinkra (i. e. the underlying adinkra 
with lines topologically the same as edges on of a hypercube). 

\setcounter{subfigure}{0}
\begin{figure}[!htbp]
   \centering
 \subfigure[]{\label{f:Chiral}\includegraphics[width=0.45\columnwidth]{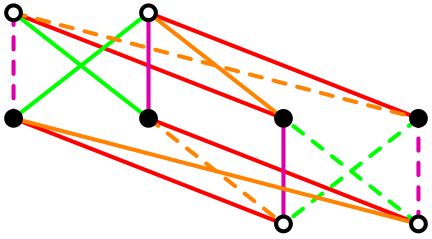}}
    \quad
    \subfigure[]{\label{f:ChiralNoPrpl}\includegraphics[width=0.45\columnwidth]{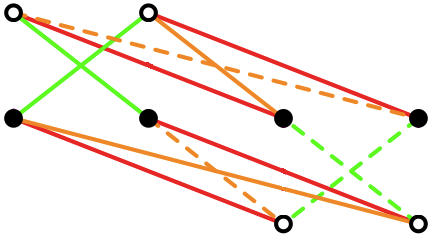}}
     \end{figure}
     \centerline{{Figure 7: A chiral supermultiplet whose associated equations can 
     lift to two}}\centerline{{dimensions (left), and a hypercubic sub-adinkra (right).}}
     
     \noindent
In Fig.\ \# 7, the adinkra is liftable by assigning ``+" to purple and green colors, 
while ``$-$" to all other colors in the adinkra. All bow ties will be unidextrous, 
thereby preserving liftability. 

We can, however, determine liftability of the folded adinkra on a subset of colored 
lines that can make the adinkra ``hypercubic", which reduces calculations, and 
simplifies the liftability algorithm because we can look at fewer lines. If we call 
one color ``non-hypercubic" in Fig.\ \# 7(a), such as the purple line, then the 
sub-adinkra made from 3 (hypercubic) lines is liftable because the green lines 
make bow ties with no other color, as seen in Fig.\ \# 7(b). We claim that 
determining lift ability of the hypercubic sub-adinkra, determines liftability 
for the entire adinkra.
  
To prove this, however, we must first prove that hypercubic lines and 
non-hypercubic lines always form bow ties with each other in an adinkra.
       \newline\newline
$\mathbf{Lemma~1:}$\newline
Non-hyperbolic lines always create bow ties with the hypercubic lines in an adinkra.
\newline\newline
$\mathbf{Proof:}$

Let us make a folded adinkra with only 1 designated non-hypercubic line color, and 
$r$ hypercubic line colors.   Because the hypercubic lines are not unique, we can 
remove some $1$ hypercubic line to re-create a hypercubic adinkra (forcing the 
non-hypercubic line to be part of a hypercubic adinkra). Because there are $r+1$ 
nodes on the second row, but only $r$ lines, the adinkra cannot be fully extended 
(i.e. must contain bow ties). If the non-hypercubic line doesn't create bow ties with 
those hypercubic lines, then we can add the previous hypercubic colored lines 
back into the adinkra, and take another hypercubic line color out. Doing this some 
finite number of times, we will always obtain non-hypercubic lines creating bow 
ties with hypercubic lines because the adinkra contains bow ties. If the non-hypercubic 
lines never created bow ties with the hypercubic lines, then we could create a fully 
extended hypercubic sub-adinkra which is impossible.

Now, let us assume this is true for a folded adinkra with $k-1$ non-hypercubic lines, 
and $r$ hypercubic lines such that $k\le\frac{r!}{3!(r-1)!}$, the total number of 
non-hypercubic lines possible. Therefore, $r+k$ nodes connect to the $r+k$ lines 
for one of the top nodes which again. By again removing any $k-1$ non-hypercubic 
lines, we find that the $k^{th}$ non-hypercubic line makes bow ties with at least one 
other hypercubic line color by a similar argument to the one non-hypercubic line case.
\begin{flushright}$\Box$\end{flushright}

Now we have the tools to prove the theorem. We will use the above lemma to show 
that, if ambidextrous bow ties always appear in hypercubic sub-adinkras, then adding 
non-hypercubic line colors will not make the adinkra liftable.
\newline\newline
$\mathbf{Theorem~1:~Hypercube~Adinkra~Liftability~Theorem}$\newline
The adinkra is liftable if and only if the adinkra's hypercubic lines contain no 
ambidextrous bow ties.
\newline\newline
$\mathbf{Proof:}$

Let us first assume the hypercubic lines contain ambidextrous bow ties. Because 
non-hypercubic lines create bow ties with hypercubic lines by Lemma 1, if the 
hypercubic lines contain ambidextrous bow ties, then by adding 1 non-hypercubic 
line, our only option to remove ambidextrous bow ties is to make the sign of all 
hypercubic lines $+1$ or $-1$. If that is true, however, then the non-hypercubic 
line will have to have the opposite sign (in which case it creates an ambidextrous 
bow tie) or same sign, which is impossible. Therefore adding 1 non-hypercubic 
line doesn't make the adinkra liftable. If we add $k$ hypercubic lines, then by 
Lemma 1, they too will create bow ties with hypercubic lines, so by a similar 
argument, these cannot make the adinkra liftable.

Now let us go the other direction and assume the adinkra is not liftable. If the 
adinkra's hypercubic lines contain ambidextrous bow ties then we're done. If, 
however, the non-hypercubic lines contain ambidextrous bow ties, then all the 
non-hypercubic line have to be the same sign to remove their non-ambidextrous 
bow ties. because every non-hypercubic line has bow ties with hypercubic lines, 
all the lines must be the same sign to remove non-ambidextrous bow ties which 
is impossible. We can remove $k$ hypercubic lines as before, and still find the 
adinkra is not liftable, because ambidextrous bow ties will still exist. Now, by the 
previous paragraph, we can add the $k$ lines back and find that the non-hypercubic 
lines also create ambidextrous bow ties. By removing the $k$ non-hypercubic 
lines, we then find that the hypercubic lines create ambidextrous bow ties, which 
finishes our proof.
\begin{flushright}$\Box$\end{flushright}

\section{Are Adinkras Deemed Liftable by Electric Shapes?}

\setcounter{subfigure}{0}
\begin{figure}[!htbp]
   \centering
    \subfigure[]{\label{f:diamondPath}
    \includegraphics[width=0.3\columnwidth]{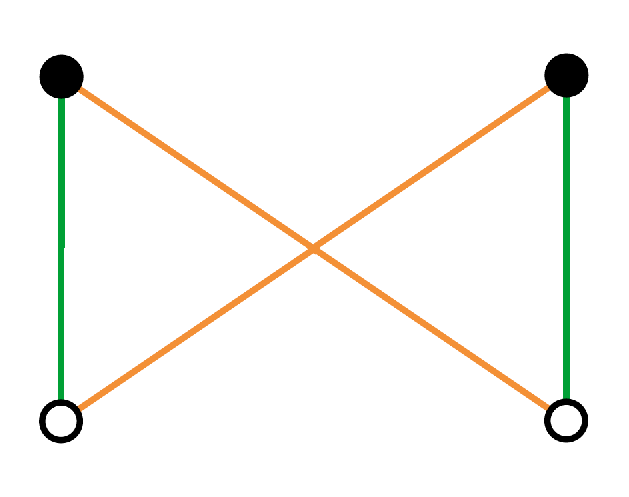}}
          \put(-97,1){$1$}      \put(-24,1){$2$}
          \put(-97,86){$1$}      \put(-24,86){$2$}
    \quad
    \subfigure[]{\label{f:bowtiePath}
    \includegraphics[width=0.3\columnwidth]{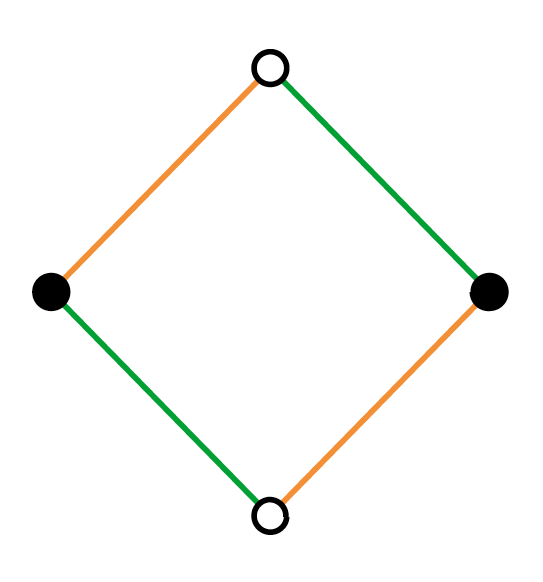}} 
          \put(-51,3){$1$}         \put(-51,112){$2$}
          \put(-93,59){$1$}      \put(-2,59){$2$}
    \label{f:diambtPath}
\end{figure}
\centerline{{Figure 8: Bow tie and diamond two-color closed paths with line dashing 
removed.}}
 \noindent

Before we introduce the algorithm, we will introduce color matrices from which we 
can determine bow ties and diamonds. 

The bow tie adinkra to the left in Fig.\ \# 8 is called a ``valise" and naturally leads 
to a lexicographic numbering of the nodes as shown (in this section we remove 
line dashing for clarity).  The diamond adinkra on the right in Fig.\ \# 8 is obtained 
from the bow tie adinkra by `lifting' the bosonic 2 node on the lower right of the 
bow tie adinkra.

Color matrices are formed by decomposing an adinkra into its monochromatic 
components, as shown in Fig.\ \# 9 for the bow tie, and in Fig.\ \# 10 for the diamond. 
It is obvious in Fig.\ \# 8 that the lines are topologically the same as the edges of 
a square. Once we've partitioned the set of line colors into $k$ non-hypercubic 
colors, and $\mathcal{N}-k$ hypercubic colors, we can introduce parameters $
\b{}_{\rI}^{\pm1}$, where ``I" corresponds to the I${}^{th}$ hypercubic line color.  We 
assign a value of $\b{}_{\rI}$ or $\b{}_{\rI}{}^{-1}$ depending  on whether the colored 
line segment is above or below the node attached to it. The $\b_i^{\pm1}$ assigning 
is equivalent to multiplying a fermion or boson field, by 1 or $\partial_\tau$ in an 
associated equation within an adinkra, depending on whether the fermion or boson's 
associated node is above or below its superpartner.

\begin{figure}[!htbp]
\centering
\includegraphics[width=0.8\columnwidth]{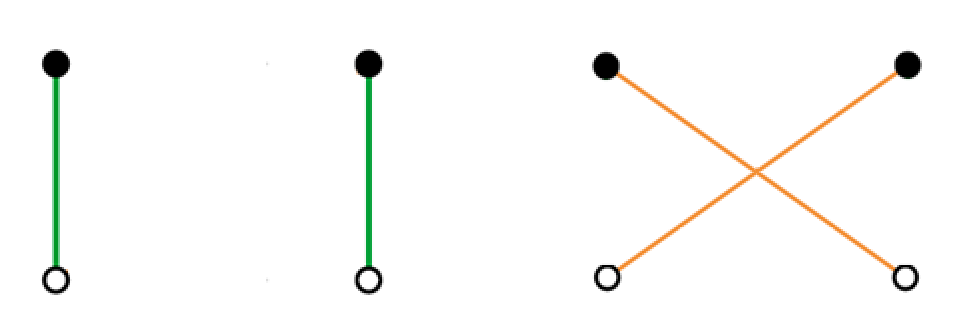}
\put(-290,102){$1$}   \put(-190,102){$2$}  \put(-114,102){$1$}   \put(-18,102){$2$}  
\put(-290,-4){$1$}      \put(-190,-4){$2$}      \put(-114,-4){$1$}      \put(-18,-4){$2$}  
\label{f:bowtieN}
\end{figure}
\centerline{{Figure 9: The $\b$-parameter assignments to bow tie edges.}}
\noindent

 The numerical labels attached to each the nodes allows us to translate each diagram
into a color matrix.  We associate each bosonic nodal label with a row entry, and each 
fermionic nodal label with a column entry in the color matrix denoted $\mathcal{B}{}_{
\rI}{}_{R}$, while we associate each fermionic node label with a row and bosonic node 
label with a column in the matrix denoted $\mathcal{B}{}_{\rI}{}_{L}$. Thus for the bow 
tie decomposition, shown in Fig.\ \#9, the green color matrices are

\be\label{eq:BowtieB}
{\mathcal{B}}_{1L}= \left( \begin{array}{cc}
\beta_1^{-1} & 0\\ 
0 & \beta_1^{-1} \\ 
\end{array} \right)~~~,~~~~ 
 \mathcal{B}_{1R}= \left( \begin{array}{cc}
\beta_1& 0\\ 
0 & \beta_1 \\ 
\end{array} \right)~~~,~~~~ 
\ee
while the yellow color matrices are
\be
\mathcal{B}_{2L}= \left( \begin{array}{cc}
0 & \beta_2^{-1}\\ 
\beta_2^{-1} & 0 \\
\end{array} \right)  ~~~,~~~~ \\
 \mathcal{B}_{2R}= \left( \begin{array}{cc}
0 & \beta_2\\ 
\beta_2 & 0 \\
\end{array} \right)  ~~~, 
\ee

\begin{figure}[!htbp]
\centering
\includegraphics[width=0.8\columnwidth]{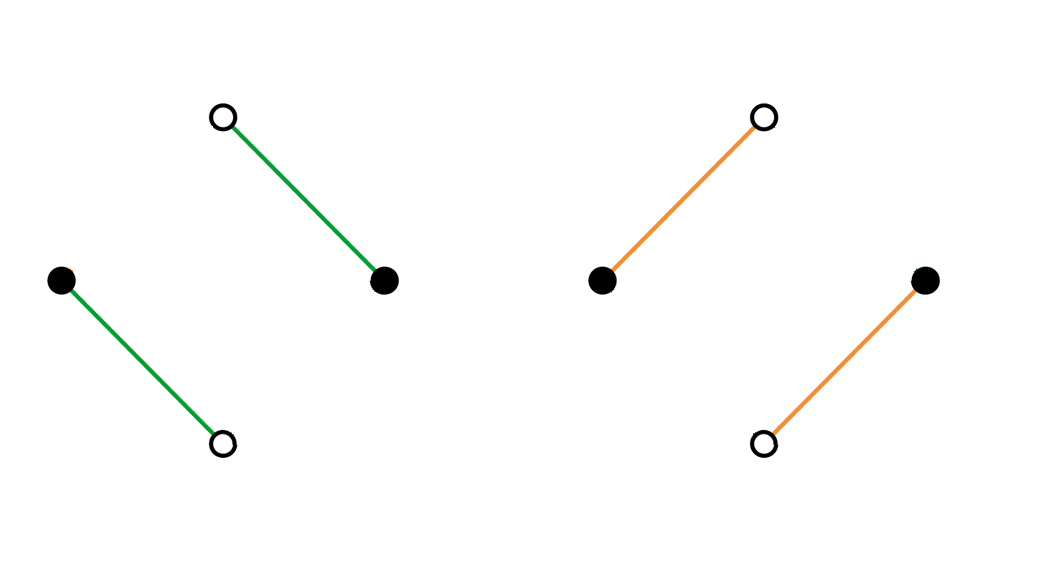}
\put(-260,124){$2$}  \put(-180,80){$2$}     \put(-74,126){$2$}   \put(-24,80){$2$} 
\put(-308,80){$1$}     \put(-218,20){$1$}    \put(-118,80){$1$}   \put(-74,18){$1$}  
\put(-308,-10){{Figure 10: The $\b$-parameter assignments to diamond edges.}}
    \label{f:diamondN}
\end{figure}
\noindent

In comparison the green lines, for the diamond adinkra's color matrices, shown in Fig.\ \#10 
below, are
\be\label{eq:DiamondB}
{\mathcal{B}}_{1L}= \left( \begin{array}{cc}
\beta_1^{-1} & 0\\ 
0 & \beta_1 \\ 
\end{array} \right)~~~,~~~~ 
\mathcal{B}_{1R}= \left( \begin{array}{cc}
\beta_1 & 0\\ 
0 & \beta_1^{-1} \\ 
\end{array} \right)~~~,~~~~ 
\ee
while, for the yellow lines, the associated matrices are
 \be
 \mathcal{B}_{2L}= \left( \begin{array}{cc}
0 & \beta_2\\ 
\beta_2^{-1}  & 0 \\
\end{array} \right)   ~~~,~~~~\\
 \mathcal{B}_{2R}= \left( \begin{array}{cc}
0 & \beta_2\\ 
\beta_2^{-1}  & 0 \\
\end{array} \right)   ~~~.
\ee
Multiplying $\mathcal{B}_{1R}\mathcal{B}_{2L}$ together with the vector of fermions, 
$\mathbf{\Psi}$, is equivalent to an automorphism along the yellow, and then green 
lines from one fermion node to the other. To make this path tracing more explicit, let 
us first define the color dependent block matrix, which permutes the super vector 
$\mathbf{\Phi}\oplus\mathbf{\Psi}$ as defined below.\newline $~$ \newline
\noindent
$\mathbf{Definition~1:}$\newline
Let the matrix $\mathbf{C{}_{{}_{\rm J}}}$ be the color block matrix associated with 
the J${}^{th}$ line color. We define 
\be
\mathbf{C{}_{{}_{\rm J}}}\equiv
\left( \begin{array}{cc}
0&\mathcal{B}{}_{{}_{\rm J}R}\\ 
 \mathcal{B}{}_{{}_{\rm J}L}&0\\ 
\end{array} \right) ~~~, 
\ee
\noindent
to represent the permutation on the vector of bosons and fermions by the operator $
D_j$. Thus, when we trace a path from one node to another via the green line, we 
are really multiplying the vector of fermions and bosons, $\mathbf{\Phi}\oplus\mathbf{
\Psi}$ by $\mathbf{C}_1$. \newline

Therefore, multiplying the matrices $\mathbf{C}_1\mathbf{C}_2$ for the bow 
For 
the bow tie in Fig.\ \# 8, we find that 
\be
\mathbf{C}_1\mathbf{C}_2=
\left( \begin{array}{cccc}
0&\b_2^{-1}\b_1&0&0\\
\b_2^{-1}\b_1&0&0&0\\
0&0&0&\b_2\b_1^{-1}\\
0&0&\b_2\b_1^{-1}&0\\
\end{array} \right) ~~~,
\ee
\noindent
in comparison, the diamond 2-color path in Fig.\ \# 8 yields
\be
\mathbf{C}_1\mathbf{C}_2=
\left( \begin{array}{cccc}
0&\b_2\b_1&0&0\\
\b_2^{-1}\b_1^{-1}&0&0&0\\
0&0&0&\b_2^{-1}\b_1\\
0&0&\b_2\b_1^{-1}&0\\
\end{array} \right)  ~~~.
\ee
tie traces 
part of the bow tie path starting from every node. Because it is the relative position of 
nodes along a path that distinguishes diamonds and bow ties, we will show that 
examining the $C_1 C_2$ block matrix will tell us which 2-color loop is which. 
The difference between these two 2-color paths becomes apparent by studying the 
matrix eigenvalues. The eigenvalues of the color block matrix are $\pm\{\b_2\b_1^{
-1},\b_2^{-1}\b_1\}$, while the diamond adinkra's eigenvalues are $\pm1$. We don't
even have to find the eigenvalues of the entire matrix, however because the upper 
matrix $\mathcal{B}{}_{{}_{\rm I}L}\mathcal{B}{}_{{}_{\rm J}R}$ already says the number 
of bow ties.  Intuitively, we can see this by remembering that the non-zero elements 
in $\mathcal{B}{}_{{}_{\rm I}L}\mathcal{B}{}_{{}_{\rm J}R}$ record the path along half 
of a 2-color loop for each boson (as there are two bosons in each 2-color loop, the 
relative positions of nodes in the entire loop is recorded). The eigenvalues record 
the relative positions of the lines as we move around the full 2-color loop. Thus, if 
we move around a diamond adinkra, we move down lines of color I and J and then 
back up another pair of lines of color I and J on our return trip. Recording the round 
trip gives us the values
\be
\pm \b{}_{{}_{\rm I}}\b{}_{{}_{\rm J}}\b{}_{{}_{\rm I}}^{-1}\b{}_{{}_{\rm J}}^{-1} = \pm 1  ~~~.
\ee
Similarly, a bow tie 2-color loop's possible eigenvalues are
\be
\pm \b{}_{{}_{\rm I}}^{-1}\b{}_{{}_{\rm J}}~,~\pm\b{}_{{}_{\rm I}}\b{}_{{}_{\rm J}}^{-1}  ~~~.
\ee 
Let us define a set S of line colors I such that I and J create bow ties for some J 
$\in$ S (if no bow ties exist for the color I, then then S = J). Then we can say an 
ambidextrous adinkra is liftable if and only if $\exists$ S' of line colors I' $\notin$ 
S such that I' and J' create bow ties for some J' $\in$ S' (where similarly if J' doesn't 
exist, S'=I'):~$\forall$ I $\in$ S and I'$\in$ S',  
\be
\text{evals}\[\mathcal{B}{}_{{}_{\rm I}R}\mathcal{B}{}_{{}_{{\rm I}'} L} \] = \pm1   ~~~.
\ee
In other words, there exists at least two sets of line colors which don't create bow 
ties with each other.

\section{Emergence of the Lorentzian World Sheet}

$~~~$
There is one aspect of this approach that likely needs to be explained in further 
detail: how does an extra dimension appear simply by adding the helicity to the 
one dimensional spinors?  This part of the story begins with the algebra of the 
D's and their realization on valise supermultiplets.  By definition, the `L-matrices' 
satisfy the conditions,
\be 
\eqalign{  
 (\,{\rm L}_\rI\,)_i{}^\hj\> (\,{\rm R}_\rJ\,)_\hj{}^k + (\,{\rm L}_\rJ\,)_i{}^\hj\>(\,{\rm 
 R}_\rI\,)_\hj{}^k & = 2  \, \delta{}_{\rI \rJ}  \,\delta{}_i{}^k~~,  \cr
 (\,{\rm R}_\rJ\,)_\hi{}^j\>(\, {\rm L}_\rI\,)_j{}^\hk + (\,{\rm R}_\rI\,)_\hi{}^j\>(\,{\rm 
 L}_\rJ\,)_j{}^\hk
  &= 2\,  \delta {}_{\rI\rJ}\,\delta {}_\hi{}^\hk~~,
}  \label{GarDNAlg1} 
  \ee
where
\be
~~~~(\,{\rm R}_\rI\,)_\hj{}^k\,\delta_{ik} = (\,{\rm L}_\rI\,)_i{}^\hk\,\delta_{\hj\hk}~~.
\label{GarDNAlg2}
\ee
For a fixed size d, there are $2^{d -1} d!$ distinct matrices that can be used.
Given a set of such L-matrices, one can introduce d bosons $\Phi_i $ ($i = 
1, \, \dots, \, d $) and d fermions $ \Psi_{\hat k}$ (${\hat k}= 1, \, \dots, \, d $)
along with $N$ superderivatives ${\rm D}{}_{{}_{\rm I}}$ (${\rm I} = 1, \, \dots, 
\, N $) that satisfy the equations
\be
{\rm D}{}_{{}_{\rm I}} \Phi_i ~=~ i \, \left( {\rm L}{}_{{}_{\rm I}}\right) {}_{i \, {\hat 
k}}  \,  \Psi_{\hat k} ~~~,~~~
{\rm D}{}_{{}_{\rm I}} \Psi_{\hat k} ~=~  \left( {\rm R}{}_{{}_{\rm I}}\right) {}_{{\hat 
k} \, i}  \, \partial_{\tau} \, \Phi_{i}  ~~.
 \label{chiD0J}
\ee
The definitions in (\ref{GarDNAlg1}), (\ref{GarDNAlg2}), and (\ref{chiD0J}) will 
ensure that these D = 1 bosons and fermions form a representation of $N
$-extended SUSY.  It is a requirement of the construction that both $\Phi_i$ and 
$ \Psi_{\hat k}$ are functions of a single time-like parameter $\tau$.  These 
equations ensure that the operator equation 
\be
{\rm D}{}_{{}_{\rm I}} \, {\rm D}{}_{{}_{\rm J}} ~+~ {\rm D}{}_{{}_{\rm J}} \, {\rm 
D}{}_{{}_{\rm I}} ~=~ i \,  2\,  \delta {}_{\rI\rJ} \, \partial_{\tau}  ~~~,
 \label{chiD0J00}
\ee
is satisfied on all bosons and fermions.

When a plus or minus helicity index is appended to the D-operator in the first
equation in (\ref{chiD0J}) that index must consistently appear on both sides.
\be
{\rm D}{}_{{}_{\rm I}} \Phi_i ~=~ i \, \left( {\rm L}{}_{{}_{\rm I}}\right) {}_{i \, {\hat k}}  
\,  \Psi_{\hat k} ~~\to ~~
{\rm D}{}_{{\pm}_{\rm I}} \Phi_i ~=~ i \, \left( {\rm L}{}_{{}_{\rm I}}\right) {}_{i \, {\hat k}}  
\,  \Psi_{\pm \, \hat k}  ~~.
 \label{chiD0J1}
\ee
So the helicity assignment to the D-operators implies a helicity assignment to the
spinors.  Note there is not any requirement that the same helicity assignment be
made for all the spinors.  Some can have plus signs while other may have minus
signs.  The choice of which numerical values of the spinorial index corresponds 
to plus and which to minus is the information carried by the operator
Spin(${\rm D}{}_{{}_{\rm I}}$) and we have
\be
{\rm {Spin}}({\rm D}{}_{{}_{\rm I}}) \Phi_i ~=~ i \, \left( {\rm L}{}_{{}_{\rm 
I}}\right) {}_{i \, {\hat k}}  \,  {\rm {Spin}}( \Psi_{\hat k})   ~~~,
\ee
since the bosons are all assumed to be scalars.

Now we can consider the effect of the appended helicity assignments on
the equation in (\ref{chiD0J00}) so that it becomes
\be
{\rm {Spin}}({\rm D}{}_{{}_{\rm I}}) \, 
{\rm {Spin}}({\rm D}{}_{{}_{\rm J}}) ~+~ 
{\rm {Spin}}({\rm D}{}_{{}_{\rm J}}) \, 
{\rm {Spin}}({\rm D}{}_{{}_{\rm I}} )~=~ i \,  2\,  \delta {}_{\rI\rJ} \, 
{\rm {Spin}}(\partial_{\tau})    ~~~.
 \label{chiD0K1}
\ee
If the spin assignments of both ${\rm D}{}_{{}_{\rm I}}$ and ${\rm D}{}_{{}_{\rm J}}$
are positive, then we can assign twice these individual helicities to ${\rm {Spin}}(
\partial_{\tau}) $ and call the new derivative $\partial_{\pp}$.  As the subscript on 
this new operator indicates, its helicity is the same as two plus signs and we `stack' 
these two signs one on top of the other in a space saving manner.  If the spin 
assignments of both ${\rm D}{}_{{}_{\rm I}}$ and ${\rm D}{}_{{}_{\rm J}}$ are 
negative, then we can assign twice these individual helicities to ${\rm {Spin}}(
\partial_{\tau}) $ and call the new derivative $\partial_{\mm}$.  Once more as the 
subscript on this new operator indicates, its helicity is the same as two minus 
signs and we `stack' these two signs one on top of the other in a space saving 
manner.  If the spin assignment of ${\rm D}{}_{{}_{\rm I}}$ is different from that of 
${\rm D}{}_{{}_{\rm J}}$ it is consistent to set the RHS of (\ref{chiD0K1}) to zero.  
So in the presence of helicity assignments to the D-operators we have
\be  \eqalign{
{\rm {Spin}}({\rm D}{}_{{}_{\rm I}})  {\rm {Spin}}({\rm D}{}_{{}_{\rm J}})  +
{\rm {Spin}}({\rm D}{}_{{}_{\rm J}})  {\rm {Spin}}({\rm D}{}_{{}_{\rm I}} ) &= i   2\,  
 \delta {}_{\rI\rJ} \,   \partial_{\pp}~ ~~;{\rm {if}} ~~
  {\rm {Spin}}({\rm D}{}_{{}_{\rm I}}) \,=\,  {\rm {Spin}}({\rm D}{}_{{}_{\rm J}})
 \,=\, \fracm 12   \cr
{\rm {Spin}}({\rm D}{}_{{}_{\rm I}})  {\rm {Spin}}({\rm D}{}_{{}_{\rm J}})  +
{\rm {Spin}}({\rm D}{}_{{}_{\rm J}})  {\rm {Spin}}({\rm D}{}_{{}_{\rm I}} ) &= i   2\,  
 \delta {}_{\rI\rJ} \,   \partial_{\mm}~ ~~;{\rm {if}} ~~
  {\rm {Spin}}({\rm D}{}_{{}_{\rm I}}) \,=\,  {\rm {Spin}}({\rm D}{}_{{}_{\rm J}})
 \,=\, -\fracm 12   \cr
{\rm {Spin}}({\rm D}{}_{{}_{\rm I}})  {\rm {Spin}}({\rm D}{}_{{}_{\rm J}})  +
{\rm {Spin}}({\rm D}{}_{{}_{\rm J}})  {\rm {Spin}}({\rm D}{}_{{}_{\rm I}} ) &= 0 ~ 
~~~~~~~~~~~;{\rm {if}} ~~
  {\rm {Spin}}({\rm D}{}_{{}_{\rm I}}) \,\ne\,  {\rm {Spin}}({\rm D}{}_{{}_{\rm J}})
~~, }
 \label{chiD0K1b}
\ee
in the three respective cases.  But there are now {\em {two}} distinct bosonic derivative 
operators $\partial_{\pp}$ and $\partial_{\mm}$.  This implies that the functions 
associated with the nodes can depend on both coordinates conjugate to these two 
derivatives.  In the case of the diamond (which corresponds to the case of the d $=$ 
2 and $N$ $=$ 2 valise but with one node lifted), we see
\newline
 $$
\vCent
 {\setlength{\unitlength}{1mm}
  \begin{picture}(-20,0)
   \put(-66,-30){\includegraphics[width=1.4in]{figures/Diamond}}     
  \put(-77,-12){$\Psi_1(\tau) $} \put(-30,-12){$\Psi_2(\tau)$}  
    \put(-60,-29){$\Phi(\tau) $} \put(-44,4){$F(\tau)$}  
  \end{picture}}
$$
 $$
\vCent
{\setlength{\unitlength}{1mm}
\begin{picture}(-20,0)
\put(-16,-10){\includegraphics[width=.8in]{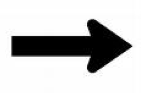}}
\end{picture}}
$$
 $$
\vCent
 {\setlength{\unitlength}{1mm}
 \begin{picture}(-160,-10)
 \put(22,-13){\includegraphics[width=1.4in]{figures/Diamond}} 
 \put(5,12){$\Psi_{+1}(\s^{\pp},  \s^{\mm}) $} \put(52,-1){$\Psi_{-2}(\s^{\pp},  
 \s^{\mm})$}  
 \put(29,-15){$X(\s^{\pp},  \s^{\mm}) $} \put(44,21){$F(\s^{\pp},  \s^{\mm})$} 
 \put(-60,-20){Figure 11: Addition of helicity and two dimensions emergence.}
  \end{picture}}
$$  \vskip0.8in \noindent
which shows the effect of adding the ambidextrous helicity.  Thus, the diamond
adinkra in 1D gives rise to the superfield containing the coordinate $X(\s^{\pp}, 
 \s^{\mm}) $ required to describe the world sheet of the superstring in its RNS
 formulation.

There is one more bonus to this construction.  The two derivatives that appear
by this construction are naturally associated with {\em {light-cone}} coordinates 
for a two dimensional manifold with a Lorentzian symmetry.   
\be
\s^{\pp} ~=~   \fracm 1{\sqrt 2}   \left[ \, \tau ~+~ \sigma \, \right] ~~~,~~~ \s^{\mm} ~=~ 
 \fracm 1{\sqrt 2}   \left[ \, \tau ~-~ \sigma \, \right] 
 ~~~~.
 \label{chiD0K3}
\ee
Spin realizing a Lorentzian (1, 1) signature (as compared to a (2, 0) or (0, 2) 
signature) is certainly {\em {not}} an inessential feature of this construction.
It is as if adinkras together with helicity conservation, for its own internal 
consistency, requires a Lorentzian structure and a second coordinate to 
appear.  Thus, the world sheet of superstring theory emerges with remarkable
simplicity in this way.

We have thus shown how to start from representations of (21) and derive 
representations of
\be
{\rm D}_{\a \, i} \, {\rm D}_{\b \, j} ~+~  {\rm D}_{\b \, j} \, {\rm D}_{\a \, i}  ~=~
i \, 2 \, \delta_{i \, j} \,  (\g^{\underline a}){}_{\a \,\b } \, \pa_{\underline a}
 ~~~,
\ee
i.\ e., the two dimensional version of supersymmetry with $(\g^{\underline a}){}_{\a \,\b } $
being 2D Dirac matrices and $\pa_{\underline a}$ = ($\pa_{\pp}$, $\pa_{\mm}$).
Some explicit examples are given in the appendices.

\section{Finding the Liftability of the Tesseract \& Folded Tesseract Adinkras}

$~~~$ In previous works \cite{Codes}, it was shown that four color adinkras have the 
topology of the tesseract, or by using self-dual block linear error correcting codes, a 
folded version of the tesseract.  Using our algorithm applied to the folded tesseract, 
we find 6 liftable supermultiplets in Appendix A. The chiral multiplets found agree 
nicely with Thm. \# 1 (where we removed one ``non-hypercubic" line to determine 
liftability). An example of a liftable adinkra is in Fig.\ \# 7. Because the liftable chiral 
supermultiplets are easy to determine, we used them to test our algorithm on folded 
adinkras. As we saw in Fig.\ \# 7, all liftable chiral adinkras were determined by 
removing a single line color, in agreement with Thm. \# 1.

In addition, we determined the liftability of tesseract adinkras, as seen in Appendix 
B (a typical adinkra is seen in Fig.\ \# 6). To determine which adinkras were liftable, 
we were forced to use Mathematica due to the shear number of adinkras possible 
(there were 90 unique adinkras, of which 22 are liftable). We wrote a code that 
looked at each adinkra, and determined liftability via the above algorithm. Interestingly, 
the shape of the liftable adinkras make up 5 distinct classes. Almost all adinkras within 
these classes differ only by placement of line color: some adinkras have red-green bow 
ties  while others have yellow-green bow ties, but all the adinkras within a class ``look" 
the same, ignoring the line color.  

As a demonstration of the comprehensive nature of the code and its power to uncover 
two dimensional supermultiplets it was used on two classes of four color adinkras:  
\newline $~~~$ (a.) adinkras with four black dots and four white dots   
where  \newline $~~~~~~~~~$ the algorithm
identified the ($p$, $q$) $=$ ($2$, $2$) chiral and 
\newline $~~~~~~~~~$ twisted chiral 
(2$|$4$|$2) supermultiplets,
\newline $~~~~~~~~~$ 
\newline $~~~$ (b.) adinkras with eight black dots and eight white dots where 
\newline $~~~~~~~~~$ 
the algorithm identified  the following ambidextrous supermultiplets:
\newline $~~~~~~~~~$ 
(1.) the ($p$, $q$) $=$ ($2$, $2$) \& ($p$, $q$) $=$ ($3$, $1$) real scalar (1$|$4$|$6$|$4$|$1) 
\newline $~~~~~~~~~~~~~\,~~$ supermultiplet, 
\newline  $~~~~~~~~~$ 
(2.) the  ($p$, $q$) $=$ ($2$, $2$) semi-chiral (2$|$6$|$6$|$2) supermultiplet, 
\newline  $~~~~~~~~~$ 
(3.) the ($p$, $q$) $=$ ($2$, $2$)  \& ($p$, $q$) $=$ ($3$, $1$) (4$|$8$|$4) supermultiplets,
and
\newline  $~~~~~~~~~$ 
(4.) the ($p$, $q$) $=$ (3, 1) of (1$|$5$|$7$|$3) and (3$|$7$|$5$|$1) supermultiplets.
\newline \noindent
Above the notation ($n_1 \, | \,n_2 \, | \,\dots$) denotes the number of nodes at each
height in the adinkra starting at the bottom, i.e. the fields of lowest engineering
dimension.  

Before we leave this discussion of liftability, comment seems warranted with regard 
to a result found in the work of Ref.\  \cite{DimUp}.  The authors analyzed the case 
of lifting from one dimension the adinkras that correspond to the usual 4D, $\cal N$ 
$=$ 1 chiral and vector supermultiplets.  The adinkras that describe the component 
fields of a chiral supermultiplet have the form shown for all of the (2$|$4$|$2) adinkras 
in appendix A.  The adinkras that describe the component fields of a vector 
supermultiplet have the (3$|$4$|$1) height form.

In our exploration of liftability, we find that the ambidextrous (3$|$4$|$1) adinkras 
{\em {cannot}} be lifted to even two dimensions while the  (1$|$4$|$6$|$4$|$1) can 
be lifted to two dimensions.  This is exactly the same as what was found in Ref.\  
\cite{DimUp}.  The (3$|$4$|$1) adinkra corresponds to a component-level Coulomb 
and Wess-Zumino gauged-fixed description of the vector supermultiplet.  Our current 
work on dimensional extension to two dimensions and the work in Ref.\  \cite{DimUp} 
on dimensional extension to four dimensions both imply that it is the 
(1$|$4$|$6$|$4$|$1) adinkra that is liftable in both cases.  

As was noted in Ref.\  \cite{DimUp}, this implies that the fields that were zero in the 
Coulomb and Wess-Zumino gauged-fixed description of a vector supermultiplet must 
be restored to allow a description compatible with adinkra lifting.  Stated in an 
alternative manner, the component fields of a vector supermultiplet must be 
embedded into a (1$|$4$|$6$|$4$|$1) adinkra in order to be lifted.  The extra fields 
were given the name `spectators' in \cite{DimUp} and these precisely correspond 
to the fields that are zero in a Wess-Zumino and Coulomb gauge.

However, our presentation emphasizes a feature that is obscured by the discussion 
in Ref.\  \cite{DimUp}.  As we find this condition already emerges in going from one 
dimension to two dimensions, this means that the spinor bundles that were introduced 
in the discussion in Ref.\  \cite{DimUp} are not critical.  It is the absence of two-color 
ambidextrous bow tie paths that is important.  

Furthermore, this is apparently the origin of the gauge symmetries associated with 
Yang-Mills superfields.  The fact that the (3$|$4$|$1) exists as a well defined adinkra 
representation in one dimension but can {\em {only be lifted to two dimensions if it 
is embedded}} into a (1$|$4$|$6$|$4$|$1) $=$ (1$|$4$|$3$|$0$|$0) + (0$|$0$|
$3$|$4$|$1) structure implies there must exist some transformations that make the 
(1$|$4$|$6$|$4$|$1) adinkra equivalent to the  (3$|$4$|$1) adinkra.  These are 
precisely the Yang-Mills type gauge transformations acting on the 
(1$|$4$|$6$|$4$|$1) adinkra.  It is the absence of bow ties that permits gauge 
superfields to exist in dimensions greater than one.

Our results also show the original of the gauge symmetries associated with the
anti-symmetric 2-form is of a similar nature.  As shown in the work of \cite{G-1},
there exist a valise formulation of the reduced tensor supermultiplet.  Being
a valise, it has the height structure of a (4$|$4) height representation which is actually
the Wess-Zumino gauge-fixed and Coulomb gauge-fixed version of a (4$|$8$|$4) 
height representation where the lowest nodes of the (4$|$8$|$4) are fermions.
This particular  (4$|$8$|$4) can be decomposed as (4$|$8$|$4) $=$ (4$|$4$|$0) 
+ (0$|$4$|$4).  Here, the  (4$|$4$|$0) is the field strength of the vector multiplet
while the (0$|$4$|$4) are the components of the tensor supermultiplet that remain
in the Wess-Zumino gauge-fixed and Coulomb gauge-fixed version of the tensor
supermultiplet.  So once again, it is the absence of bow ties that permits gauge 
superfields to exist in dimensions greater than one.

We thus find the results that of the three distinct adinkras containing four 
bosonic nodes and four fermionic nodes obtained in \cite{G-1} to be lifted
on a world sheet containing ($p$, $q$) $=$ ($2$, $2$) supersymmetry,
these must occur as (2$|$4$|$2) representations, or contained in (1$|$4$|$6$|$4$|$1)
representations, or within  (4$|$8$|$4) representations.  As the condition
of ($p$, $q$) $=$ ($2$, $2$) supersymmetry is a precursor for a full
4D, $\cal N$ = 1 theory, these are precisely the configurations needed
to obtain the full 4D, $\cal N$ = 1 chiral, vector, and tensor supermultiplets
respectively.

\section{Conclusions}

$~~~$
It has been our position \cite{ENUF}, that traditional and by now standard approaches 
to understanding the representation theory of off-shell supersymmetry
in {\em {all}} dimensions beyond one leave enormous room for additional development.  
This was the primary reason in 2001 we launched a program to tackle such problems.  
The results of this search have been the discovery of previously unknown connections 
to mathematical structures that have been hidden by the traditional approaches to 
supersymmetrical theories.

However, the greatest benefits to possessing a clean and concise understanding 
of the mathematical origins of supersymmetrical representation theory go beyond 
the purely theoretical and academic understanding of the subject.  There are 
practical benefits to be obtained.  

The first such benefit for higher dimensional
supersymmetry multiplets occurred in the work of Ref.\ \cite{N2}.  In this past work, 
a new off-shell representation of 4D, $\cal N$ $=$ 2 supersymmetry was presented.
Though much more study on this must be done, it is possible this will give new 
insight into the off-shell structure of the standard $\cal N$ $=$ 2 hypermultiplet. 
Although the methods used to present this supermultiplet were not manifestly 
based on adinkras, it was adinkra-based concepts utilized by one of the authors 
of the work that gave an {\em a} {\em {priori}} indication that such a higher 
dimensional supermultiplet existed.  

In the current presentation, included in the appendices, we give evidence for the
discovery of a new 2D,  $\cal N$ $=$ 2 supermultiplet with 8 - 8 components.
Should this interpretation receive wider validation, it will provide further vindication 
for initiating this approach to understanding off-shell supersymmetrical representation
theory and show the benefits of being able to systematically scan the zoo of 1D 
adinkras to find those that are liftable to 2D.  Let us also note that this new representation
is the Klein flipped version of the chiral spinor superfield reduced to two dimensions.
Once more as this is not a representation that has been discussed previously, we
see the utility of a systematic search of the adinkra zoo.

We have shown that the state-of-the-art in the program has come to the point where 
an algorithm can be written to search for two dimensional off-shell supermultiplets.  
Our scan of the 1D adinkra zoo has been limited, but there are no {\em a} {\em 
{priori}} reasons it could not be extended to a broader search for more two 
dimensional supersymmetric representations.   We know of no other approaches 
that efficiently and practically allow for such a search to be undertaken. \newline
$~$  \newline
$~$  \newline
$~$

$$~~$$
\noindent
{\Large\bf Added Note In Proof}
After the completion of this work, we were reminded of the research by
James Park \cite{JPap} which also explores the liftability of bowtie and
diamond adinkras, but utilizing the compatibility conditions described
in the work in \cite{DimUp}.

${~~~}$ \newline \noindent
${~~~~~}$``{\it {The soul never thinks without a picture.
}}"${~~~}$ \newline
\newline $~~~~~~~$ -- Aristotle
$$~~$$
\noindent
{\Large\bf Acknowledgments}

This research was supported in part by the endowment of the John S.~Toll 
Professorship, the University of Maryland Center for String \& Particle Theory, 
National Science Foundation Grant PHY-09-68854.  KB would like to thank 
Thomas Rimlinger for determining liftable adinkras by hand, in order to check 
over the Mathematica code.  Additionally KB acknowledges participation in 
2012 SSTPRS (Student Summer Theoretical Physics Research Session).  
Adinkras were drawn with \emph{Adinkramat} $\copyright$ 2008 by 
G. Landweber.

\newpage
\noindent
{\Large {\bf {Appendix A: Liftable 2D Folded-Tesseract Based $~$ }}} \vskip.005in
 {\Large {\bf {$~~~~~~$  $~~$  $~\,~~~$ Adinkras \& Associated 2D SUSY 
  $~~$ }}} \vskip.005in
  {\Large {\bf {$~~\,~$  $~~~$  $~~~~~~$ Equations}}}  \vskip.03in

 Here we present liftable (2$|$4$|$2) ambidextrous supermultiplet adinkras 
 and associated 2D SUSY equations. Below are the liftable adinkras. 
$$
\vCent
{\setlength{\unitlength}{1mm}
\begin{picture}(-26,0)
\put(-67,-76){\includegraphics[width=5.2in]{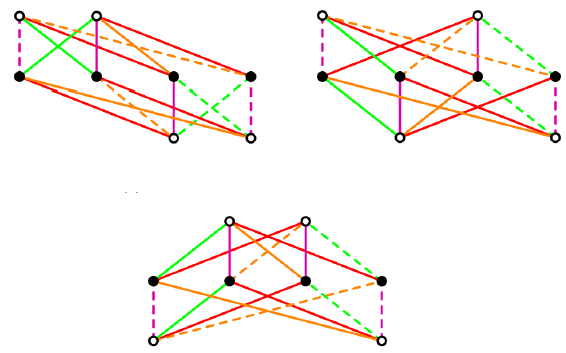}}
\put(-42,-40){(a)}
\put(36,-40){(b)}
\put(-8,-80){(c)}
\put(-60,-88){Figure 12: The liftable $(2|4|2)$ folded-tesseract based adinkras}
\end{picture}}
 \nonumber
$$ \vskip3.4in

Let us define green, orange, red and purple line colors to be associated with the
the subscripts ``1", ``2", ``3", and ``4", respectively. We label open (boson)
 nodes as $\Phi_i$, where the ``$i$" subscript lexicographically labels nodes along 
 each row from left to right, where the top left open node is $\Phi_1$. Closed nodes 
 corresponding to fermion fields are labeled $ \Psi_{\hat k} $, where the ``${\hat k}$" 
 subscript is organized analogously.  The SUSY equations for (a) are:
$$
\vCent
{\setlength{\unitlength}{1mm}
\begin{picture}(-26,0)
\put(-78,-58){\includegraphics[width=1\columnwidth]{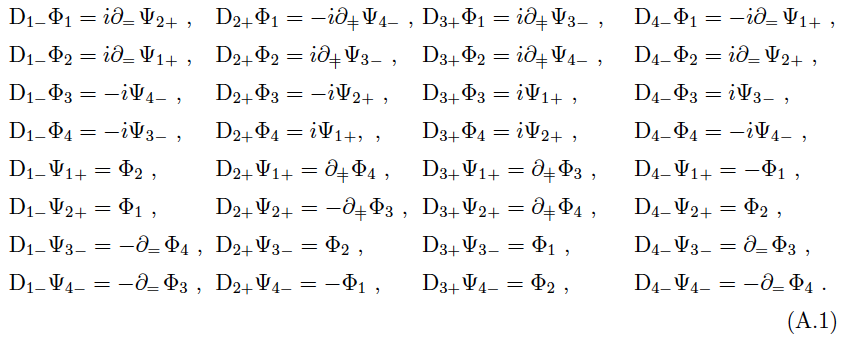}}
  \end{picture}}
 \nonumber
$$ \vskip2in \noindent
The SUSY equations for (b) are equivalent with Red $\rightarrow$ ${\rm 
D}_1$, Green $\rightarrow$ ${\rm D}_3$ and Purple $\rightarrow -D_4$. All other 
colors are defined as they were originally.   Similarly, the SUSY equations for (c) 
are equivalent with Yellow $\rightarrow D_1$, Red $\rightarrow$ ${\rm D}_2$, and 
Green $\rightarrow$ ${\rm D}_3$.
$~~$ \newline
$~~$ \newline
$~~$ \newline
\noindent
{\Large {\bf {Appendix B: Liftable 2D Tesseract  Based Adinkras }}}  \vskip.005in
{\Large {\bf {$~~~$  $~~~~~$  $~~~~$ \& Associated 2D SUSY Equations
  $~~$ }}}

In this appendix we present twenty-two liftable ambidextrous adinkras, organized by 
number of nodes in each row. Below each set, we will present in complete detail a 
representative of the associated 2D SUSY equations.  There are also given a set of
redefinition give that allow the construction of  the 2D SUSY equations for the other 
members in each set.   

Our initial definitions are such that we define green lines to correspond to the D-operator 
with the subscript ``1", purple lines to correspond to the D-operator the subscript ``2", 
red lines to correspond to the D-operator the subscript ``3", and yellow lines to
correspond to the D-operator the subscript ``4". We label open (boson) nodes as $
\Phi_i$, where the ``$i$" subscript lexicographically labels nodes along each row from 
left to right, where the top left open node is $\Phi_1$. Closed nodes corresponding to 
fermion fields are labeled as $\Psi_{\hat k}$, where the ``${\hat k}$" subscript is 
organized analogously.  Many adinkras create equivalent equations except with colors 
re-defined. In this appendix, all colors not explicitly redefined, remain as they were 
originally defined above.

The first liftable ambidextrous adinkra tesseract-based is shown in Fig.\ \# 13 below.
$$
\vCent
{\setlength{\unitlength}{1mm}
\begin{picture}(-26,0)
\put(-56,-48){\includegraphics[width=4in]{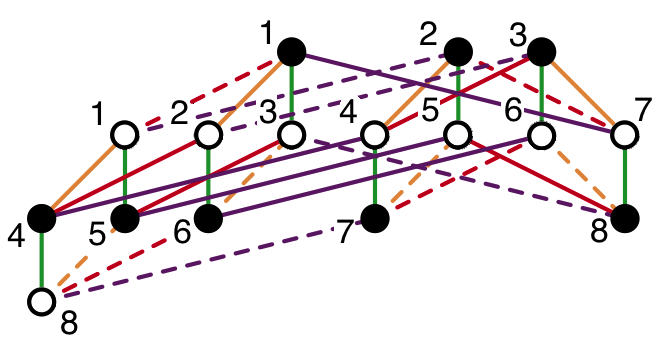}}
\put(-58,-53){Figure 13: A liftable $(1|5|7|3)$ adinkra with nodal labels shown.}
  \end{picture}}
 \nonumber
$$ \vskip2in \noindent
This adinkra may be engineered to produce a ($p$, $q$)  = (3, 1) superfield.
We will use this example to show the nodal labels in the context of the adinkra,
although we will not relate field labels to the associated nodes in any of our
later demonstrations in this appendix, these can easily be restored by using
this as an example of the process.
$$
\vCent
{\setlength{\unitlength}{1mm}
\begin{picture}(-26,0)
\put(-79,-109){\includegraphics[width=6.5in]{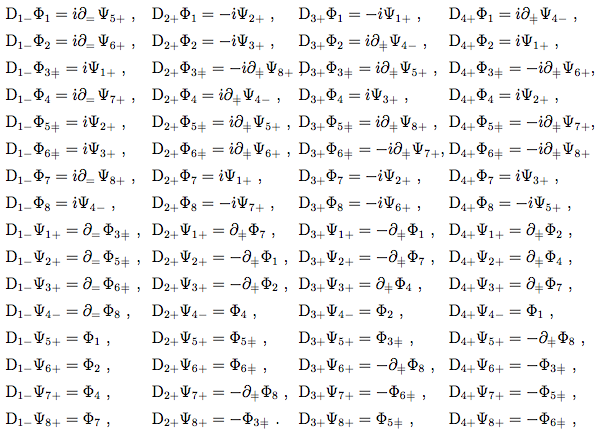}}
\put(66,-112){$({\rm B.1}) $}
  \end{picture}}
 \nonumber
$$
\vskip4in     
 $~~$ \newline  \noindent
The bosons have spin 0 (e.g. $\Phi_1$, $\Phi_2$, $\Phi_4$, $\Phi_7$, $\Phi_8$) or 
correspond to the positive helicity components of a spin 1 vector (e.g. $\Phi_3{}_\pp$, 
$\Phi_5{}_\pp$, $\Phi_6{}_\pp$).  We also note that $\Psi_4{}_-$ represents a fermion 
with spin -1/2 and all other fermions have spin $1/2$. 

The adinkra shown in Fig.\ \# 14, actually leads to two distinct superfields.  One 
\vskip0.05in
$$
\vCent
{\setlength{\unitlength}{1mm}
\begin{picture}(-26,0)
\put(-38,-51){\includegraphics[width=3.in]{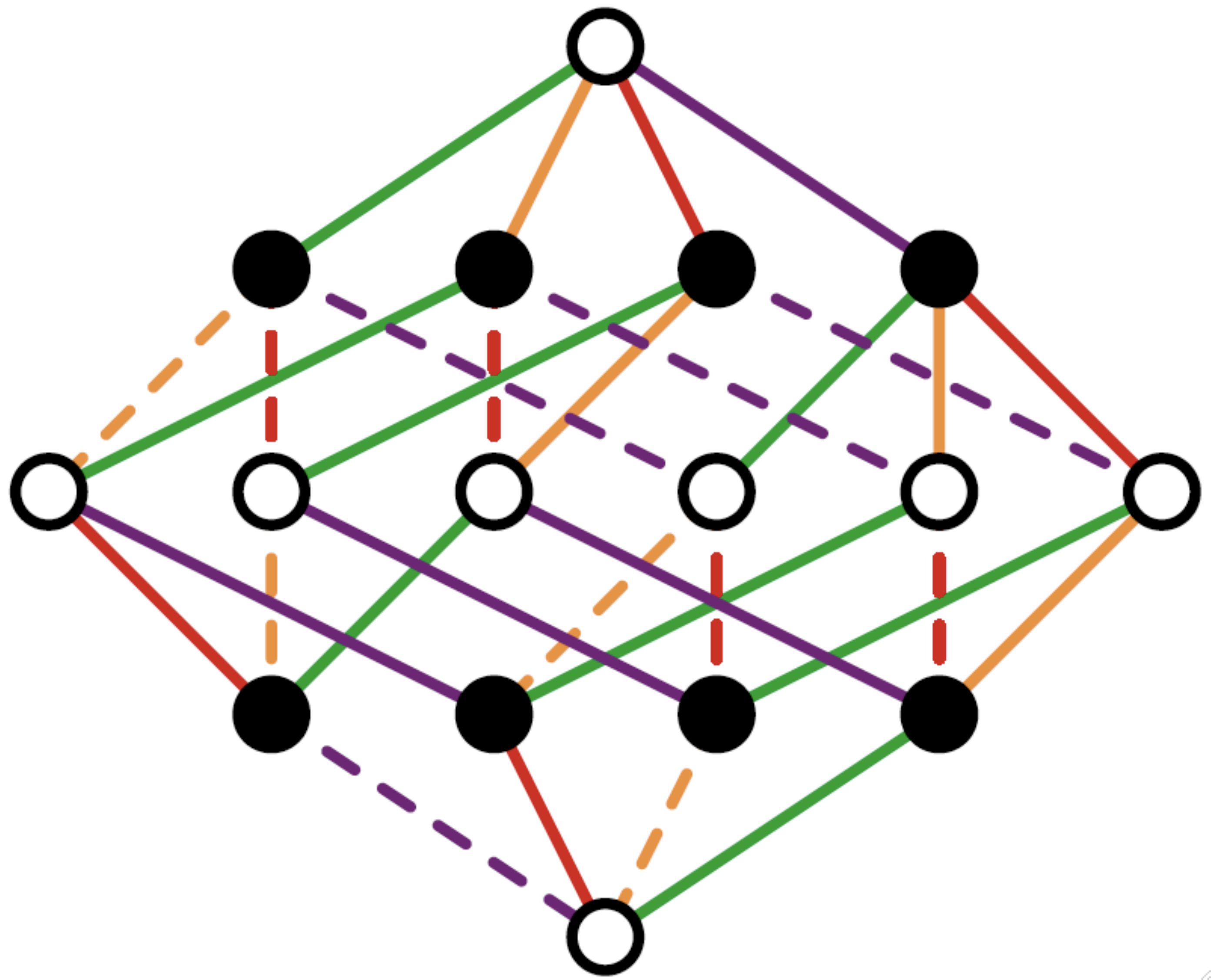}}
\put(-55,-57){Figure 14: The unique (1$|$4$|$6$|$4$|$1) fully extended adinkra.}
\end{picture}}
\nonumber
$$ \vskip2.2in  \noindent
of these corresponds to the dimensional reduction of the well known real scalar 
superfield from 4D, $\cal N$ $=$ 1 supersymmetry. Accordingly, it has a degree 
of extended supersymmetry characterized by ($p$, $q$) $=$ (2, 2).  As this is a 
familiar representation, we will not discuss it further.  

However, there is a second superfield characterized by ($p$, $q$) $=$ (3, 1) and 
as this is the lesser know supersymmetric representation we will concentrate on 
its structure\footnote{Of course there is also the possibility of ($p$, $q$) $=$ (1, 3) 
but this is simply the parity \newline $~~~~~$ reflection of the  ($p$, $q$) $=$ 
(3, 1) superfield.}.  The associated 2D SUSY equations (containing a spin 3/2 
fermion $\Psi_8{}_{\mm}^+$) are: 
$$
\vCent
{\setlength{\unitlength}{1mm}
\begin{picture}(-26,0)
\put(-79,-109){\includegraphics[width=6.05in]{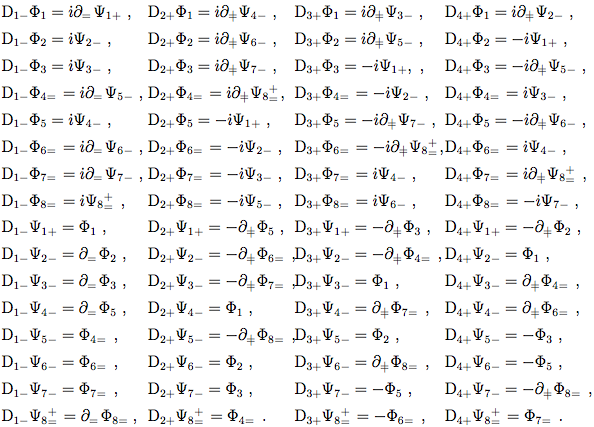}}
\put(66,-112){$({\rm B.2}) $}
  \end{picture}}
 \nonumber
$$ \vskip4.2in
 For the next set of four graphs shown in Fig.\ \# 15 and 
 for which we will explicitly present a
 representative set of equations correspond to:  
\vskip0.05in
$$
\vCent
 {\setlength{\unitlength}{1mm}
  \begin{picture}(-26,0)
   \put(-76,-25){\includegraphics[width=6.in]{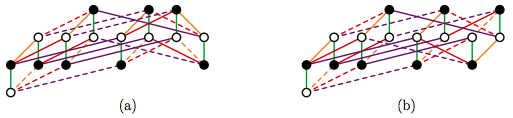}}
  \end{picture}}
 \nonumber
$$   \noindent \vskip.6in
$~$
  \vskip3.5in \noindent 
\newpage

$$
\vCent
{\setlength{\unitlength}{1mm}
\begin{picture}(-26,0)
\put(-75,-30){\includegraphics[width=2.5in]{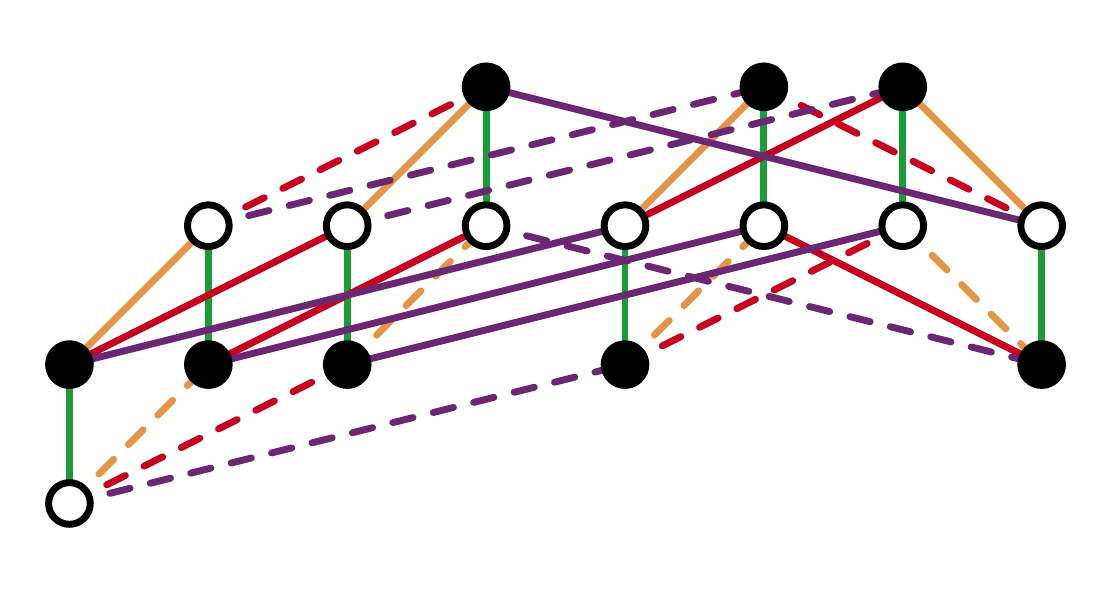}}
\put(-40,-30){(a)}
\end{picture}}
\nonumber
\vCent
{\setlength{\unitlength}{1mm}
\begin{picture}(-26,0)
\put(-77,-31){\includegraphics[width=6.in]{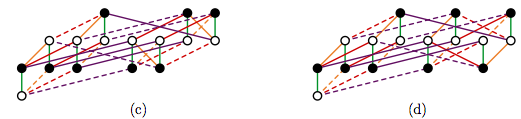}}
\put(-66,-38){Figure 15: All adinkras of the form (1$|$5$|$7$|$3)
that lift to two dimensions.}
\end{picture}}
 \nonumber
$$ \vskip1.5in
For Fig. \# 14(a), the associated SUSY equations are:
$$
\vCent
{\setlength{\unitlength}{1mm}
\begin{picture}(-26,0)
\put(-78,-109){\includegraphics[width=6.05in]{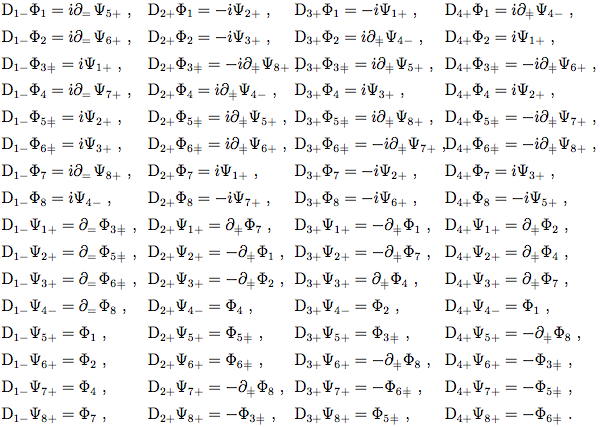}}
\put(66,-112){$({\rm B.3}) $}
  \end{picture}}
 \nonumber
$$ \vskip4.2in
The other adinkras have equivalent equations with color redefinitions. To create the 
corresponding SUSY equations for the other adinkras in Fig. \# 14, we first redefine 
colors according to: \newline
$~~$ (b). Green $\rightarrow$ ${\rm D}_4$  $~~~~~~~~~~~~~~~~$(c.) Green 
$\rightarrow$ ${\rm D}_4$ $~~~~~~~~~~~~~~~~$(d.)  Green $\rightarrow$ ${\rm D}_3$

$~~~$ Yellow $\rightarrow$ ${\rm D}_1$   $~~~~~~~~~~~~~~~~~~~~$ Yellow 
$\rightarrow$ ${\rm D}_3$ $~~~~~~~~~~~~~~~~~~~~~$Yellow $\rightarrow$ 
${\rm D}_4$

$~~~~~~~~~~~~~~~~~~~~~~~~~~~~~~~~~~~~~~~~~~~$ Red $\rightarrow$ 
${\rm D}_1$  $~~~~~~~~~~~~~~~~~~~~~~~~$Red $\rightarrow$ ${\rm D}_2$

$~~~~~~~~~~~~~~~~~~~~~~~~~~~~~~~~~~~~~~~~~~~~~~~~~~~~~~~~~~$
$~~~~~~~~~~~~~~~~~~~~~~~~$ Purple $\rightarrow$ ${\rm D}_1$
\newpage \noindent
So that we see the need to redefine only two colors in case (b), three in case (c) 
and all four in case (d). The corresponding equations are not the exact same, 
however, due to differences in line dashing. We may, however, re-create the 
exact same equations with node redefinitions, (e.g. $\Phi_i \rightarrow -\Phi_i$) 
in the cases of (b) and (d).  We do note, however, if one set of equation has an 
odd number of line dashing relative to another, the two sets of equation describe 
a supermultiplet and its twisted version\cite{G-1}.

Turning to the adinkras in Fig.\ \# 16 shown below. We find six cases of adinkras 
that can lifted.  They can be used to engineer ($p$, $q$) = (2, 2) semi-chiral 
superfields \cite{SeMiCH1}, \cite{SeMiCH2}, and \cite{SeMiCH3}.  
$$
\vCent
{\setlength{\unitlength}{1mm}
\begin{picture}(-26,0)
\put(-78,-136){\includegraphics[width=5.5in]{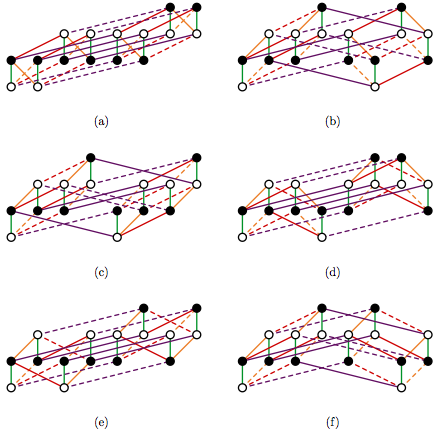}}
\put(-70,-143){Figure 16: All adinkras of the form (2$|$6$|$6$|$2) that lift to two
dimensions.}
  \end{picture}}
 \nonumber
$$ \vskip5.4in \noindent
The associated equations for the adinkra in Fig. \# 16(a) are:
$$
\vCent
{\setlength{\unitlength}{1mm}
\begin{picture}(-26,0)
\put(-78,-8){\includegraphics[width=6.05in]{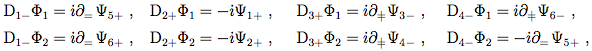}}
\end{picture}}
\nonumber
$$ 
\newpage
$$
\vCent
{\setlength{\unitlength}{1mm}
\begin{picture}(-26,0)
\put(-78,-90){\includegraphics[width=6.05in]{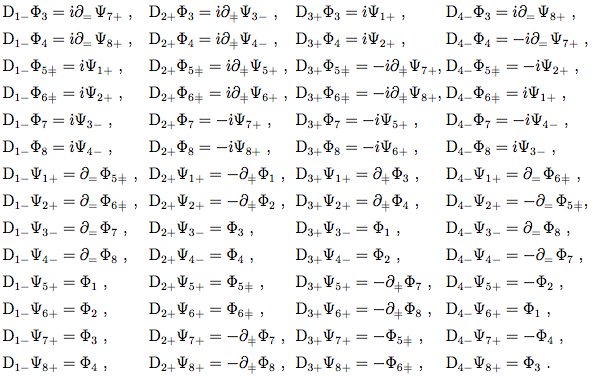}}
\put(66,-96){$({\rm B.4}) $}
 \end{picture}}$$ \vskip3.6in \noindent
To create equivalent equations for the other adinkras in Fig. \# 16 from those in
 Fig. \# 16 (a), we re-defining the colors as follows:

(b). Green $\rightarrow$ ${\rm D}_1$  $~~~~~~~~~~~~~~~~~~~~$(c.) Green 
$\rightarrow$ ${\rm D}_3$ $~~~~~~~~~~~~~~~$(d.)  Green $\rightarrow$ $
{\rm D}_1$

$~~~~~$ Yellow $\rightarrow$ ${\rm D}_3$  $~~~~~~~~~~~~~~~~~~~~~~~~~$ 
Yellow $\rightarrow$ ${\rm D}_4$$~~~~~~~~~~~~~~~~~~~~~$Yellow $
\rightarrow$ ${\rm D}_2$

 $~~~~~$ Red $\rightarrow$ ${\rm D}_4$  $~~~~~~~~~~~~~~~~~~~~~~~~~~~~~$ 
 Red $\rightarrow$ ${\rm D}_1$  $~~~~~~~~~~~~~~~~~~~~~~~$Red $\rightarrow$ 
 ${\rm D}_3$
  
$~~~~~~~~~~~~~~~~~~~~~~~~~~~~~~~~~~~~~~~~~~~~~~~~~~~~~~~~~$
$~~~~~~~~~~~~~~~~~~~~~~~~~~~~~~~~~~~~~~$Purple $\rightarrow$ ${\rm D}_4$

(e.)  Green $\rightarrow$ ${\rm D}_3$ $~~~~~~~~~~~~~~~~~~~~$(f.)  Green 
$\rightarrow$ ${\rm D}_2$

$~~~~~$ Yellow $\rightarrow$ ${\rm D}_1$  $~~~~~~~~~~~~~~~~~~~~~~$
$~~$Yellow $\rightarrow$ ${\rm D}_3$

$~~~~~$  Red $\rightarrow$ ${\rm D}_2$ $~~~~~~~~~~~~~~~~~~~~~~~~~~~$ 
Red $\rightarrow$ ${\rm D}_4$

$~~~~~$   Purple $\rightarrow$ ${\rm D}_4$$~~~~~~~~~~~~~~~~~~~~~~~~~
~$Purple $\rightarrow$ ${\rm D}_1$

The graphs in Fig.\ \# 17 below provide adinkras as a basis to engineer ($p$, 
$q$) = (3, 1) supermultiplets.  The associated equations from adinkra in Fig. \# 
16(a) are:
$$
\vCent
{\setlength{\unitlength}{1mm}
\begin{picture}(-26,0)
\put(-78,-22){\includegraphics[width=6.05in]{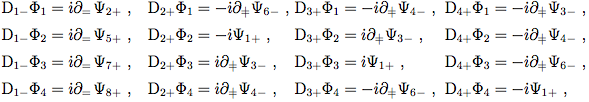}}
\end{picture}}
\nonumber
$$
\newpage

$$
\vCent
{\setlength{\unitlength}{1mm}
\begin{picture}(-26,0)
\put(-78,-76){\includegraphics[width=6.05in]{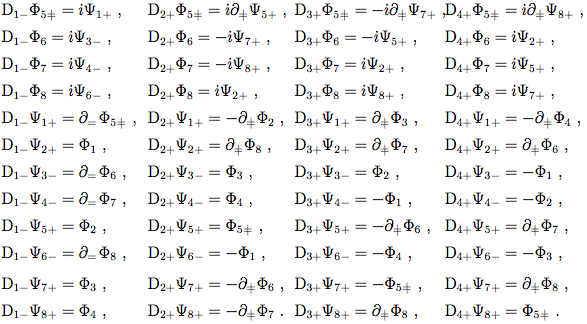}}
\put(66,-80){$({\rm B.5}) $}
 \end{picture}}$$ \vskip2.9in
To create equivalent equations for the rest of the adinkras in Fig.\ \# 17 from 
the equations in Fig.\ \# 17(a), we redefine colors below for the other adinkras:

(b). Green $\rightarrow$ ${\rm D}_4$$~~~~~~~~~~~~~~$(c.) Green $\rightarrow$ 
${\rm D}_4$$~~~~~~~~~~~~~~$(d.)  Green $\rightarrow$ ${\rm D}_3$

$~~~~~$ Yellow $\rightarrow$ ${\rm D}_1$$~~~~~~~~~~~~~~~~~$$~~$ Yellow 
$\rightarrow$ ${\rm D}_3$ $~~~~~~~~~~~~~~~~$$~~$Yellow $\rightarrow$ $
{\rm D}_4$

$~~~~~~~~~~~~~~~~~~~~~~~~~~~~~~~~~~~~~~~~~~~$ Red $\rightarrow$ $
{\rm D}_1$$~~~~~~~~~~~~~~~~~~~~~~$Red $\rightarrow$ ${\rm D}_2$

$~~~~~~~~~~~~~~~~~~~~~~~~~~~~~~~~~~~~~~~~~~~~~~~~~~~~~~~~~$
$~~~~~~~~~~~~~~~~~~~~~~$ Purple $\rightarrow$ ${\rm D}_1$

$$
\vCent
{\setlength{\unitlength}{1mm}
\begin{picture}(-26,0)
\put(-78,-36){\includegraphics[width=5.5in]{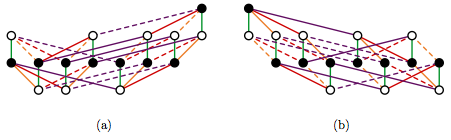}}
\put(-78,-79){\includegraphics[width=5.5in]{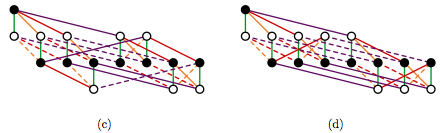}}
\put(-35,-83){Figure 17: Four (3$|$7$|$5$|$1) adinkras.}
  \end{picture}}.
 \nonumber
$$ 
\vskip2.2in
\newpage

There are seven adinkras in our next set of adinkras for engineering. However, 
they split into two categories: (a)-(d) may be used to create ($p$, $q$) =  (3, 1) 
superfields, while (e)-(g) may be used to create ($p$, $q$) =  (2, 2) superfields.  
These are all shown in Fig.\ \# 18 below.
$$
\vCent
{\setlength{\unitlength}{1mm}
\begin{picture}(-26,0)
\put(-66,-133){\includegraphics[width=5.20in]{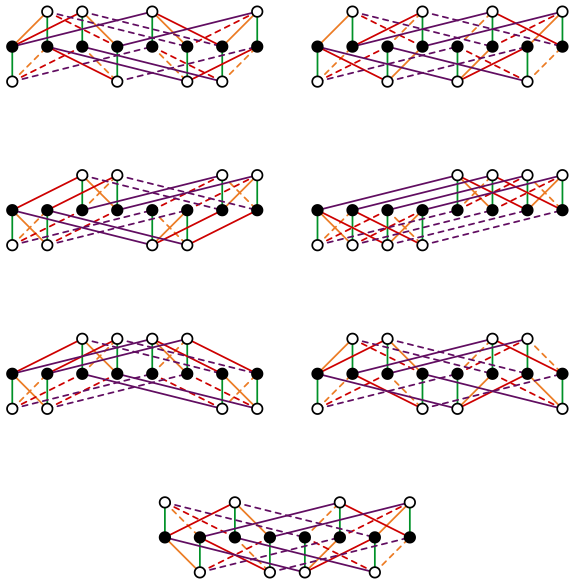}}
\put(-40,-28){(a)}  \put(34,-28){(b)}
\put(-40,-64){(c)}  \put(34,-64){(d)}
\put(-40,-104){(e)}  \put(34,-104){(f)}
\put(0,-140){(g)}
\put(-65,-148){Figure 18: All adinkras of the form (4$|$8$|$4) that lift to two 
dimensions.}
\end{picture}}$$
\newline \noindent  \vskip5.25in
The associated equations for the adinkra in Fig \# 18(a) are:
$$
\vCent
{\setlength{\unitlength}{1mm}
\begin{picture}(-26,0)
\put(-78,-37){\includegraphics[width=6.05in]{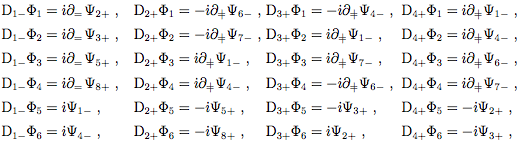}}
 \end{picture}}$$

\newpage
$$
\vCent
{\setlength{\unitlength}{1mm}
\begin{picture}(-26,0)
\put(-78,-58){\includegraphics[width=6.05in]{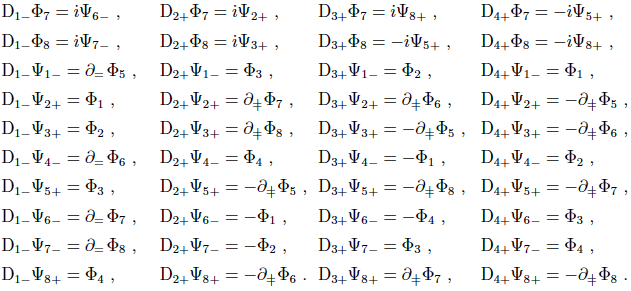}}
\put(66,-63){$({\rm B.6}) $}
\end{picture}}$$ \vskip2.3in
Line color redefinitions below of the adinkras in Fig \# 18 (b) through (d) create 
equivalent equations:

(b). Green $\rightarrow$ ${\rm D}_4$$~~~~~~~~~~~~~~~$(c.) Green $\rightarrow$ 
${\rm D}_4$$~~~~~~~~~~~~~~~~~~$(d.)  Green $\rightarrow$ ${\rm D}_4$

$~~~~~$ Yellow $\rightarrow$ ${\rm D}_1$$~~~~~~~~~~~~~~~~~~~$ Yellow $
\rightarrow$ ${\rm D}_3$ $~~~~~~~~~~~~~~~~~~~~~$$~~$Yellow $\rightarrow$ 
${\rm D}_2$

$~~~~~~~~~~~~~~~~~~~~~~~~~~~~~~~~~~~~~~~~~~~$ Red $\rightarrow$ ${\rm 
D}_1$$~~~~~~~~~~~~~~~~~~~~~~~~~~~$Red $\rightarrow$ ${\rm D}_3$

$~~~~~~~~~~~~~~~~~~~~~~~~~~~~~~~~~~~~~~~~~~~~~~~~~~~~~~~~~~~~$ 
$~~~~~~~~~~~~~~~~~~~~~~~~$ Purple $\rightarrow$ ${\rm D}_1$

It is also very important to note that (e) - (g) correspond to ($p$, $q$) = (2, 2) superfields.
The adinkra in Fig \# 18(e) has the associated equations,
$$
\vCent
{\setlength{\unitlength}{1mm}
\begin{picture}(-26,0)
\put(-78,-89){\includegraphics[width=6.05in]{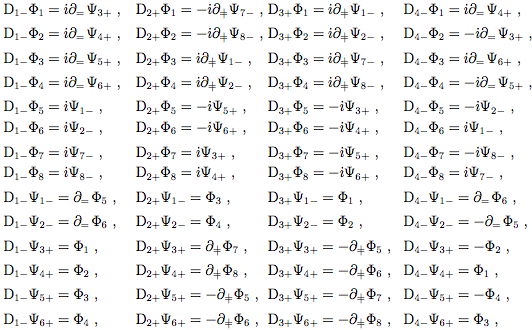}}
 \end{picture}}$$

 \newpage
 $$
\vCent
{\setlength{\unitlength}{1mm}
\begin{picture}(-26,0)
\put(-78,-2){\includegraphics[width=6.05in]{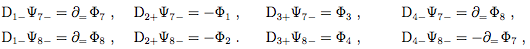}}
\put(66,-7){$({\rm B.7}) $}
 \end{picture}}$$ \vskip.1in \noindent
Similarly, redefining the colors below creates equivalent equations for the adinkras in 
Fig. \# 19(f.) and (g.):

(f.) Green $\rightarrow$ ${\rm D}_3$$~~~~~~~~~~~~~~~~~~~~$(g.)  Green $\rightarrow$ 
${\rm D}_1$

$~~~~~$ Yellow $\rightarrow$ ${\rm D}_1$$~~~~~~~~~~~~~~~~~~~~~~~~~$Yellow $
\rightarrow$ ${\rm D}_3$

$~~~~~$ Red $\rightarrow$ ${\rm D}_4$$~~~~~~~~~~~~~~~~~~~~~~~~~~~~~$Red $
\rightarrow$ ${\rm D}_4$
$~$ \newline
$$~~$$
\noindent
{\Large {\bf {Appendix C: The ($p$, $q$) = ($2$, $2$)  (4$|$8$|$4) 2D Super-}}}  \vskip.005in
{\Large {\bf {$~~~$  $~~~~$  $~~$ $~~$ multiplet  Is Not Two (2$|$4$|$2) 2D
  $~~$ }}}   \vskip.005in
  {\Large {\bf {$~~~$  $~~~$  $~~$ $~~~$ Supermultiplets 
  $~~$ }}}

In this appendix, we wish to examine more closely the ($p$, $q$) = ($2$, $2$) 
(4$|$8$|$4) 2D supermultiplet defined by (B.7).  For this purpose, let us 
re-examine a subset of the equations given by
$$ 
\eqalign{
{\rm D}_{1-} \Phi_5 &= \dot{\imath} \Psi_1{}_-   ~~,~~
 {\rm D}_{2+} \Phi_5 = -\dot{\imath} \Psi_5{}_+ ~~,~~
 {\rm D}_{3+}\Phi_5=-\dot{\imath}\Psi_3{}_+ ~~,~~~ 
 {\rm D}_{4-} \Phi_5=-\dot{\imath}\Psi_2{}_-  ~~,~~ \cr
 {\rm D}_{1-} \Phi_6 &= \dot{\imath} \Psi_2{}_-  ~~,~~ 
 {\rm D}_{2+} \Phi_6 = - \dot{\imath} \Psi_6{}_+  ~~,~~ \,
 {\rm D}_{3+}\Phi_6=-\dot{\imath}\Psi_4{}_+ ~,~~ 
 {\rm D}_{4-} \Phi_6= \dot{\imath}\Psi_1{}_-  ~~ ,~~ \cr
{\rm D}_{1-} \Phi_7 &=\dot{\imath}  \Psi_7{}_-  ~~,~~  
{\rm D}_{2+} \Phi_7 =\dot{\imath} \Psi_3{}_+  ~~,~~~~\,
{\rm D}_{3+} \Phi_7 = -\dot{\imath}\Psi_5{}_+ ~~,~~ 
{\rm D}_{4-}\Phi_7=- \dot{\imath}\Psi_8{}_- ~~,~~  \cr 
{\rm D}_{1-} \Phi_8 &= \dot{\imath} \Psi_8{}_-  ~~,~~
{\rm D}_{2+} \Phi_8 =\dot{\imath}\Psi_4{}_+  ~~,~~ ~~
{\rm D}_{3+} \Phi_8 = - \dot{\imath}\Psi_6{}_+ ~~,~~ \,
{\rm D}_{4-}\Phi_8=\dot{\imath}\Psi_7{}_-  ~~,~~
}
\eqno(C.1)
$$ 
and eliminate all the fermions of negative helicity from these equations
to obtain
$$
{\rm D}_{1-}  \left[  \begin{array}{c}
~\Phi_5  \\
~\Phi_6\\
~\Phi_7 \\
~\Phi_8 \\
\end{array}\right] ~=~  {\rm D}_{4-}  \left[  \begin{array}{c}
~\Phi_6  \\
-\, \Phi_5\\
~\Phi_8 \\
-\, \Phi_7 \\
\end{array}\right]  ~~~.
\eqno(C.2)
$$
Of course, we could also choose to eliminate the positive helicity 
fermions from (C.1) to obtain
$$
{\rm D}_{3+}  \left[  \begin{array}{c}
~\Phi_5  \\
~\Phi_6\\
~\Phi_7 \\
~\Phi_8 \\
\end{array}\right] ~=~   {\rm D}_{2+}  \left[  \begin{array}{c}
-\, \Phi_7  \\
-\, \Phi_8\\
~\Phi_5 \\
~\Phi_6 \\
\end{array}\right]  ~~~.
\eqno(C.3)
$$

These equations can be written more simply by introducing a four component 
vector defined by
$$
{\vec {\cal A}} ~=~ \left[\begin{array}{c}
~ \Phi_5 \\
~ \Phi_6 \\
~ \Phi_7 \\
~ \Phi_8 \\
\end{array}\right]  ~~~, 
\eqno(C.4)
$$
which permits the results in (C.2) and (C.3) to be re-cast into the
forms
$$
{\rm D}_{1-} {\vec {\cal A}} ~=~  i \,   \left(  {\BldmTH  { {\rm I} \otimes \s^2 }  }  \right) \, 
 {\rm D}_{4-}   {\vec {\cal A}}
~~~,~~~ 
{\rm D}_{3+} {\vec {\cal A}} ~=~ - i  \,  \left(  {\BldmTH  { \s^2  \otimes {\rm I} }  }  \right) \, 
 {\rm D}_{2+} {\vec {\cal A}}
~~,
\eqno(C.5)
$$
or equivalently
$$
{\rm D}_{4-} {\vec {\cal A}} ~=~ - i \,   \left(  {\BldmTH  { {\rm I} \otimes \s^2 }  }  \right) \, 
 {\rm D}_{1-}   {\vec {\cal A}}
~~~,~~~ 
{\rm D}_{2+} {\vec {\cal A}} ~=~  i  \,  \left(  {\BldmTH  { \s^2  \otimes {\rm I} }  }  \right) \, 
 {\rm D}_{3+} {\vec {\cal A}}
~~.
\eqno(C.6)
$$
where $ \left(  {\BldmTH  { {\rm I} \otimes \s^2 }  }  \right) $ and $ \left(  {\BldmTH  { \s^2  
\otimes {\rm I} }  }  \right)$ are 4 $\times$ 4 matrices written in terms of the outer 
product of 2 $\times$ 2 matrices.

The equations in (C.5) and (C.6) are completely consistent with the ($p$, $q$) = 
($2$, $2$) supersymmetry algebra.  However, the real issue is whether the multiplet
described by (B.6) can be equivalent to {\em {two}} copies of the multiplets described
in (A.1).  We believe the following argument show this is not possible.

Construct a quantity denoted by ${\vec {\cal B}}$ which is composed of two copies (an 
$A$ copy and a $B$ copy) of the fields associated with the lowest nodes in Fig.\ \# 12 
of the form
$$
{\vec {\cal B}} ~=~   \left[\begin{array}{cccc}
m_{11} & m_{12} &m_{13} & m_{14}  \\
m_{21} & m_{22} &m_{23} & m_{24}  \\
m_{31} & m_{32} &m_{33} & m_{34}  \\
m_{41} & m_{42} &m_{43} & m_{44}  \\
\end{array}\right] \,
\left[\begin{array}{c}
~ \Phi_3^{(A)} \\
~ \Phi_4^{(A)} \\
~ \Phi_3^{(B)} \\
~ \Phi_4^{(B)} \\
\end{array}\right]  ~=~  {\BldmTH {\cal M}} \, \left[\begin{array}{c}
~ \Phi_3^{(A)} \\
~ \Phi_4^{(A)} \\
~ \Phi_3^{(B)} \\
~ \Phi_4^{(B)} \\
\end{array}\right]  
\eqno(C.7)
$$
which introduces sixteen parameters $m_{11}\, \dots m_{44}$.  If the (4$|$8$|$4) supermultiplet
is a linear combination of two (2$|$4$|$2) mutlplets, then it should be possible to find a choice of
the sixteen parameters such that ${\vec {\cal B}}$ satisfies the conditions in (C.5).  In principle,
this should be quite easy as there are only eight equations implied by the conditions in (C.5).

There are only two rows in (A.1) that are relevant for this analysis and for convenience we
reproduce them below.  For completeness of our analysis there is one other matter we must 
attend.  We have established \cite{G-1} that if one begins with an adinkra for a chiral multiplet, the 
adinkra for the twisted chiral multiplet is obtained by reversing signs for an odd number of the 
D-operators that appear in the D-equations.  Picking this to be the D${}_{3+}$ operator 
we can write upon eliminating the fermions between these equations 
$$
\vCent
 {\setlength{\unitlength}{1mm}
 \begin{picture}(-20,0)
 \put(-56,-2){$ {\rm D}_{1-} \Phi_3^{(A)} = {\rm D}_{4-} \Phi_4^{(A)}  ~~~~~ ,$}
 \put(-5,-2) {${\rm D}_{2+} \Phi_3^{(A)}  =-  \,  \xi^{(A)}  \, {\rm D}_{3+}\Phi_4^{(A)}  ,$}
 \put(-56,-9){$ {\rm D}_{1-} \Phi_4^{(A)}  = -  {\rm D}_{4-} \Phi_3^{(A)} ,$} 
 \put(-5,-9) {${\rm D}_{2+} \Phi_4^{(A)}  =   \,  \xi^{(A)}  \, {\rm D}_{3+}\Phi_3^{(A)}  ,$}
 \put(66,-5){$({\rm C.8}) $}
\end{picture}}
$$
 \vskip.2in \noindent
where the parameter $\xi^{(A)}$ satisfies  $[\xi^{(A)}]^2$ = 1.  For the chiral multiplet we
choose $\xi^{(A)}$ = 1 for the chiral multiplet and  $\xi^{(A)}$ = $-1$  for the twisted
chiral multiplet.  We now take (C.7) and  (C.8) and check to see if the conditions in (C.5)
can be satisfied by the sixteen parameters.

For the first condition in (C.5) we find
$$ \eqalign{
{\rm D}_{1 -} {\vec {\cal B}} &=  {\BldmTH  {\cal M}} \, {\rm D}_{1 -} 
\left[\begin{array}{c}
  \Phi_3^{(A)} \\
   \Phi_4^{(A)} \\
  \Phi_3^{(B)} \\
  \Phi_4^{(B)} \\
\end{array}\right]       
 =  {\BldmTH  {\cal M}} \, {\rm D}_{4 -} 
\left[\begin{array}{c}
  \Phi_4^{(A)} \\
-\,    \Phi_3^{(A)} \\
   \Phi_4^{(B)} \\
-\,   \Phi_3^{(B)} \\
\end{array}\right]   =   {\BldmTH  {\cal M}}  \left(  {\BldmTH  {  {\rm I}    \otimes i  \s^2}  }  
\right)  {\rm D}_{4 -}  \left[\begin{array}{c}
 \Phi_3^{(A)} \\
   \Phi_4^{(A)} \\
  \Phi_3^{(B)} \\
  \Phi_4^{(B)} \\
\end{array}\right]  ~~.
}  \eqno(C.9)
$$
Upon comparing the last formula for the result in (C.9) with the first result of (C.5), it is 
clear that $\BldmTH {\cal M}$ must be chosen as the identity matrix.  Using the fact that 
$\BldmTH {\cal M}$ is the identity matrix we now calculate the second condition
in (C.5) to find
$$ \eqalign{
{~~~~}  {\rm D}_{2 +} {\vec {\cal B}} &=   \, {\rm D}_{2 +} 
\left[\begin{array}{c}
  \Phi_3^{(A)} \\
   \Phi_4^{(A)} \\
  \Phi_3^{(B)} \\
  \Phi_4^{(B)} \\
\end{array}\right]       
 =   \, {\rm D}_{3 +} 
\left[\begin{array}{c}
-\,    \xi^{(A)}  \Phi_4^{(A)} \\
 \xi^{(A)}   \Phi_3^{(A)} \\
-\,  \xi^{(B)}    \Phi_4^{(B)} \\
\xi^{(B)}    \Phi_3^{(B)} \\
\end{array}\right]   \cr 
&= - \fracm 12 \,    \xi^{(A)}  \left(  {\BldmTH  { ( {\rm I} + \s^3 )   \otimes i  \s^2}  }  
\right)  {\rm D}_{3 +}  \left[\begin{array}{c}
 \Phi_3^{(A)} \\
   \Phi_4^{(A)} \\
  \Phi_3^{(B)} \\
  \Phi_4^{(B)} \\
\end{array}\right]  \cr 
&{~~}~-~ \fracm 12 \,   \xi^{(B)}   \left(  {\BldmTH  {  ( {\rm I} - \s^3  )   \otimes i  \s^2}  }  
\right)  {\rm D}_{3 +}  \left[\begin{array}{c}
 \Phi_3^{(A)} \\
   \Phi_4^{(A)} \\
  \Phi_3^{(B)} \\
  \Phi_4^{(B)} \\
\end{array}\right]  \cr
&= - \fracm 12 \,  \left[  \xi^{(A)}  \left(  {\BldmTH  { ( {\rm I} + \s^3 )   \otimes i  \s^2}  }  
\right)  ~+~  \xi^{(B)}   \left(  {\BldmTH  {  ( {\rm I} - \s^3  )   \otimes i  \s^2}  }  
\right) \, \right]  {\rm D}_{3 +} {\vec {\cal B}}  ~~~.
}  \eqno(C.10)
$$
It is manifest that for no choice of the $\xi$-parameters is it possible to satisfy the
second condition in (C.5).

We thus conclude that the ($p$, $q$) = ($2$, $2$) (4$|$8$|$4) 2D supermultiplet is
a previously unobserved representation.  Should this interpretation receive wider
validation, this discovery will provide further vindication for the program begun
in 2001 and show the benefits of creating an algorithm that can systematically
scan the zoo of 1D adinkras to find those that are liftable to 2D.
$~~$ \newline $~~$ \newline 

{\Large {\bf {Appendix D: Mathematica Code for Scanning}}}  \vskip.005in
{\Large {\bf {$~~~~~~$  $~~~~$  $~~$ $~~$ 2D Liftability of 1D Adinkras
  $~~$ }}}  

Here we present the code used to determine liftable
adinkras that have been presented in this work.
We determine liftability using the valise $\mathcal{N}
= 4$ adinkra found in AdinkraMat v.1.1.  All calculations
assume we start from that adinkra and raise or lower
nodes from the positions shown in it. Nodes are labeled 
from left to right in the valise, just like in previous appendices.
Thus, the far left fermion is labeled $\psi_1$, and the far left 
boson is labeled $\phi_1$.

\noindent$
\newline
\newline \noindent \beta 1=2; \newline\beta 2=3;\newline\text{$\beta $3}=5;\newline\text{$\beta $4}=7;\newline
\text{(* We used numbers instead of keeping these in symbolic notation} \text{ to allow the com-
}\newline \text{puter to check equality (i.e. ``==") much faster than otherwise.  These are prime
}\newline\text{numbers in order to differentiate $\beta_I \beta_J $ from $\beta_K \beta_L$ *)} \newline
\newline\text{TotalAdinkraNum}=247;
\newline\text{(*roughly the number of adinkras the code looks though*)}
\newline \newline
\text{PsiDown}=0;\newline\text{PsiUp}=1;\newline\text{PhiBottom}=-1;\newline\text{PhiMid}=0;\newline\text{PhiTop}=1;\newline\text{GORP}=\text{ConstantArray}[0,\{7,5,8,\text{TotalAdinkraNum}\}];
 \newline \newline
\text{Evals}=\text{ConstantArray}[0,\{6,\text{TotalAdinkraNum}\}];\newline\text{LiftableAdinkraNum}=\text{ConstantArray}[0,54];\newline
\text{(*This array is used to find eigenvalues of the color matrices of every adinkra.}\newline
\text{``54" is the number of (non-unique) liftable adinkras this code finds*)}
 \newline \newline
\text{RedOrange}=1;\newline\text{OrangeRed}=\text{RedOrange};\newline\text{RedGreen}=2;\newline\text{GreenRed}=\text{RedGreen};\newline\text{RedPurple}=3;\newline\text{PurpleRed}=\text{RedPurple};\newline\text{OrangeGreen}=4;\newline\text{GreenOrange}=\text{OrangeGreen};\newline\text{OrangePurple}=5;\newline\text{PurpleOrange}=\text{OrangePurple};\newline\text{GreenPurple}=6;\newline\text{PurpleGreen}=\text{GreenPurple};\newline
\text{(*These are used when recording an adinkra's associated}\newline
 \text{color matrix eigenvalues, to make the indices easier to read*)}\newline
\newline
\text{AdinkraNum}=0;\newline\text{(*Counts the number of adinkras the code has so far seen}\newline
\text{ within the For loops below*)}\newline
\text{bottomnodeDown}=0;\newline
\text{bottomnodeUp}=1;\newline
\text{topnodeUp}=0;\newline
\text{topnodeDown}=1;\newline
\text{numLiftable}=0;\text{(*Counts the number of liftable}\newline 
\text{adinkras the code has seen in the For loops*)}
$
\newline
$\text{For}[\text{$\psi $1}=\text{PsiDown},\text{$\psi $1}\leq  \text{PsiUp},\text{$\psi $1}\text{++}$,
\newline $\text{For}[\text{$\psi $2}=\text{PsiDown},\text{$\psi $2}\leq  \text{PsiUp},\text{$\psi $2}\text{++}$,\newline $\text{For}[\text{$\psi $3}=\text{PsiDown},\text{$\psi $3}\leq  \text{PsiUp},\text{$\psi $3}\text{++}$,\newline $\text{For}[\text{$\psi $4}=\text{PsiDown},\text{$\psi $4}\leq  \text{PsiUp},\text{$\psi $4}\text{++}$,\newline \newline
\text{(*}\text{all} \text{the} \text{rest} \text{stay} \text{put}/\text{attached} \text{to} \text{$\phi $1}\text{*)} \newline \newline
$\text{For}[\text{$\psi $5}=\text{PsiDown},\text{$\psi $5}\leq  \text{PsiUp},\text{$\psi $5}\text{++}$,\newline \newline
 \text{(*move only 1 level*)}$ \newline \newline
 \text{For}[\text{$\psi $6}=\text{PsiDown},\text{$\psi $6}\leq  \text{PsiUp},\text{$\psi $6}\text{++}$,\newline $\text{For}[\text{$\psi $7}=\text{PsiDown},\text{$\psi $7}\leq  \text{PsiUp},\text{$\psi $7}\text{++}$,\newline $\text{For}[\text{$\psi $8}=\text{PsiDown},\text{$\psi $8}\leq  \text{PsiUp},\text{$\psi $8}\text{++}$,\newline $\text{For}[\text{$\phi $1}=\text{PhiBottom},\text{$\phi $1}\leq \text{PhiTop},\text{$\phi $1}\text{++}$,\newline \newline
\text{(*}\text{$\phi $8} \text{stays} \text{put}, \text{only} \text{$\phi $1} \text{goes} \text{up}\text{*)} \newline \newline
$\text{For}[\text{$\phi $2}=\text{PhiBottom},\text{$\phi $2}\leq  \text{PhiTop},\text{$\phi $2}\text{++}$,\newline $\text{For}[\text{$\phi $3}=\text{PhiBottom},\text{$\phi $3}\leq  \text{PhiTop},\text{$\phi $3}\text{++}$,\newline $\text{For}[\text{$\phi $4}=\text{PhiBottom},\text{$\phi $4}\leq  \text{PhiTop},\text{$\phi $4}\text{++}$,\newline $\text{For}[\text{$\phi $5}=\text{PhiBottom},\text{$\phi $5}\leq  \text{PhiTop},\text{$\phi $5}\text{++}$,\newline $\text{For}[\text{$\phi $6}=\text{PhiBottom},\text{$\phi $6}\leq  \text{PhiTop},\text{$\phi $6}\text{++}$,\newline $\text{For}[\text{$\phi $7}=\text{PhiBottom},\text{$\phi $7}\leq  \text{PhiTop},\text{$\phi $7}\text{++}$,\newline $\text{For}[\text{$\phi $8}=\text{PhiBottom},\text{$\phi $8}\leq  \text{PhiTop},\text{$\phi $8}\text{++}$,\newline 
\newline \noindent
(*This is equivalent to using Einstein notation for clarity)
\newline \noindent  \newline \noindent
$
\text{For}[\psi_I=\text{PsiDown}, \psi_I \leq  \text{PsiUp},\psi_I\text{++},\newline \noindent
\text{For}[\phi_J=\text{PhiBottom},\phi_J \leq  \text{PhiTop},\phi_J\text{++},
$
\newline \newline
(*
Basically we use many for loops to either raise the fermion nodes, which are on the bottom row, denoted as $\psi$up, or we keep them on the bottom row ($\psi$down). Similarly, we can lower a boson node ($\phi$down), keep it on the same row ($\phi$Mid) or raise this node if that is allowed ($\phi$up). Knowing when it is valise or not is determined by a massive if statement made at the beginning of the for loop:
*)\newline\newline
$
\text{(*Labels to make the If statement readable*)}
\newline\newline\text{$\psi $Middle}=(\text{$\psi $1}==\text{PsiDown}~~\&\&~~ \text{$\psi $2}==\text{PsiDown}~~\&\&~~ \text{$\psi $3}==\text{PsiDown}~~\&\&~~ \newline \text{$\psi $4}==\text{PsiDown}~~\&\&~~ \text{$\psi $5}==\text{PsiDown}~~\&\&~~ \text{$\psi $6}==\text{PsiDown}~~\&\&~~  \newline\newline
\text{$\psi $7}==\text{PsiDown}~~\&\&~~ \text{$\psi $8}==\text{PsiDown});\newline \text{NoPhiOnTop}=(\text{$\phi $1}\neq  \text{PhiTop}~~\&\&~~\text{$\phi $2}\neq \text{PhiTop}~~\&\&~~\text{$\phi $3}\neq \text{PhiTop}~~\&\&~~
 \newline
\text{$\phi $4}\neq \text{PhiTop}~~\&\&~~\text{$\phi $5}\neq \text{PhiTop}~~\&\&~~\text{$\phi $6}\neq \text{PhiTop}~~\&\&~~\text{$\phi $7}\neq \text{PhiTop}~~\&\&~~  \newline
\text{$\phi $8}\neq \text{PhiTop});\newline  \newline
\text{$\psi $all}=\text{$\psi $1}+ \text{$\psi $2}+ \text{$\psi $3}+\text{$\psi $4}+\text{$\psi $5}+ \text{$\psi $6}+ \text{$\psi $7}+\text{$\psi $8};\newline \newline \text{all$\phi $}=\text{$\phi $1}+ \text{$\phi $2}+ \text{$\phi $3}+\text{$\phi $4}+\text{$\phi $5}+ \text{$\phi $6}+ \text{$\phi $7}+\text{$\phi $8};\newline \newline \text{$\phi $1$\psi $s}=(\text{$\psi $1}+\text{$\psi $2}+\text{$\psi $3}+\text{$\psi $5});\newline \text{$\phi $2$\psi $s}=(\text{$\psi $1}+\text{$\psi $2}+\text{$\psi $4}+\text{$\psi $6});\newline \text{$\phi $3$\psi $s}=(\text{$\psi $1}+\text{$\psi $3}+\text{$\psi $4}+\text{$\psi $7});\newline \text{$\phi $4$\psi $s}=(\text{$\psi $2}+\text{$\psi $3}+\text{$\psi $4}+\text{$\psi $8});\newline \text{$\phi $5$\psi $s}=(\text{$\psi $1}+\text{$\psi $5}+\text{$\psi $6}+\text{$\psi $7});\newline\text{$\phi $6$\psi $s}=(\text{$\psi $2}+\text{$\psi $5}+\text{$\psi $6}+\text{$\psi $8});\newline\text{$\phi $7$\psi $s}=(\text{$\psi $3}+\text{$\psi $5}+\text{$\psi $7}+\text{$\psi $8});\newline\text{$\phi $8$\psi $s}=(\text{$\psi $4}+\text{$\psi $6}+\text{$\psi $7}+\text{$\psi $8});
$
\newline\newline
(*Impossible cases:\newline
	 1) if lowest boson is on the bottom but the fermion nodes above it
move\newline
	 2) if the highest boson is on the top but the fermion nodes below it
move\newline

Due to the symmetry of the adinkras, we need to look at a small fraction of the total number of raised and lowered nodes.
*)\newline\newline
$
\text{If}[
\newline
~~~(\text{$\phi $1}==\text{PhiBottom}~~\&\&~~\text{$\psi $Middle}~~\&\&~~\text{NoPhiOnTop})\newline
~~\text{(*This is equivalent to}\newline \text{keeping the first boson on the bottom row, and moving the other fermions} \newline\text{above or below a single row of fermions.}\newline \text{In other words we look into the liftability of $(1|8|7)$}\newline\text{ adinkras, $(2|8|6)$ adinkras, etc.  up to $(4|8|4)$ adinkras*)}
\newline\newline
~~~\|(\newline
~~~~~~(\text{(*1 boson down*)}
\newline
~~~~~~~~~\text{(*Here we keep the a boson node, $\phi_i$,}\newline
~~~~~~~~~\text{ on the bottom, while raising and lowering fermion nodes*)}\newline
~~~~~~~~~(\text{$\phi $1$\psi $s}==4*\text{PsiDown}~~\&\&~~\text{$\phi $1}==\text{PhiBottom}
\newline
~~~~~~~~~\&\&~~\text{NoPhiOnTop}~~\&\&~~\text{all$\phi $}==1*\text{PhiBottom})
\newline
~~~~~~~~~\|(\text{$\phi $2$\psi $s}==4*\text{PsiDown}~~\&\&~~\text{$\phi $2}==\text{PhiBottom}
\newline
~~~~~~~~~\&\&~~\text{NoPhiOnTop}~~\&\&~~\text{all$\phi $}==1*\text{PhiBottom})
\newline
~~~~~~~~~\|(\text{$\phi $3$\psi $s}==4*\text{PsiDown}~~~\&\&~~\text{$\phi $3}==\text{PhiBottom}
\newline
~~~~~~~~~~~~\&\&~~\text{NoPhiOnTop}~~\&\&~~\text{all$\phi $}==1*\text{PhiBottom})
\newline
~~~~~~~~~\|(\text{$\phi $4$\psi $s}==4*\text{PsiDown}~~~\&\&~~\text{$\phi $4}==\text{PhiBottom}
\newline
~~~~~~~~~~~~\&\&~~\text{NoPhiOnTop}~~\&\&~~\text{all$\phi $}==1*\text{PhiBottom})
\newline
~~~~~~)\newline
~~~~~~\text{(*In this case, at least one fermion is raised*)}\newline
~~~~~~\&\&~~\text{$\psi $all}\geq 1*\text{PsiUp})
\newline
\newline
~~~\|\newline
~~~(\newline
~~~~~~~(\newline
~~~~~~~~~~(\text{$\phi $5$\psi $s}==4*\text{PsiDown}~~\&\&~~\text{$\phi $5}==\text{PhiBottom}\newline
~~~~~~~~~~\&\&~~\text{NoPhiOnTop}~~\&\&~~\text{all$\phi $}==1*\text{PhiBottom})
\newline
~~~~~~~~~~\|(\text{$\phi $6$\psi$s}==4*\text{PsiDown}~~\&\&~~\text{$\phi $6}==\text{PhiBottom}\newline
~~~~~~~~~~\&\&~~\text{NoPhiOnTop}~~\&\&~~\text{all$\phi $}==1*\text{PhiBottom})
\newline
~~~~~~~~~~\|(\text{$\phi $7$\psi $s}==4*\text{PsiDown}~~\&\&~~\text{$\phi $7}==\text{PhiBottom}\newline
~~~~~~~~~~\&\&~~\text{NoPhiOnTop}~~\&\&~~\text{all$\phi $}==1*\text{PhiBottom})
\newline
~~~~~~~~~~\|(\text{$\phi $8$\psi $s}==4*\text{PsiDown}~~\&\&~~\text{$\phi $8}==\text{PhiBottom}\newline
~~~~~~~~~~\&\&~~\text{NoPhiOnTop}~~\&\&~~\text{all$\phi $}==1*\text{PhiBottom})
\newline
~~~~~~)\newline
~~~~~~\text{(*In this case, to avoid repeating adinkras,}\newline 
~~~~~~\text{we have at least 2 fermion nodes raised*)}\newline
~~~~~~\&\&~~(\text{$\psi $all}\geq 2*\text{PsiUp})
\newline
~~~)
\newline
\newline
~~~\|\newline
~~~(\newline
~~~~~~(\newline
~~~~~~~~~(\newline
~~~~~~~~~~~~\text{(*2 bosons down*)}
\newline
~~~~~~~~~~~~(\newline
~~~~~~~~~~~~\text{(*We have exactly 2 boson nodes down,}\newline
~~~~~~~~~~~~\text{ and 2 fermion nodes up. All possible combinations are below*)}\newline
~~~~~~~~~~~~~~~(\text{$\phi $1$\psi $s}==4*\text{PsiDown}~~\&\&~~\text{$\phi $1}==\text{PhiBottom})
\newline
~~~~~~~~~~~~~~~\|(\text{$\phi $2$\psi $s}==4*\text{PsiDown}~~\&\&~~\text{$\phi $2}==\text{PhiBottom})\newline
~~~~~~~~~~~~)\newline
~~~~~~~~~~~~\&\&\newline
~~~~~~~~~~~~(\newline
~~~~~~~~~~~~~~~\text{  }(\text{$\phi $3$\psi $s}==4*\text{PsiDown}~~\&\&~~\text{$\phi $3}==\text{PhiBottom})\newline
~~~~~~~~~~~~~~~\|(\text{$\phi $4$\psi $s}==4*\text{PsiDown}~~\&\&~~\text{$\phi $4}==\text{PhiBottom})\newline
~~~~~~~~~~~~~~~\|(\text{$\phi $5$\psi $s}==4*\text{PsiDown}~~\&\&~~\text{$\phi $5}==\text{PhiBottom})\newline
~~~~~~~~~~~~~~~\|(\text{$\phi $6$\psi $s}==4*\text{PsiDown}~~\&\&~~\text{$\phi $6}==\text{PhiBottom})\newline
~~~~~~~~~~~~~~~\|(\text{$\phi $7$\psi $s}==4*\text{PsiDown}~~\&\&~~\text{$\phi $7}==\text{PhiBottom})\newline
~~~~~~~~~~~~~~~\|(\text{$\phi $8$\psi $s}==4*\text{PsiDown}~~\&\&~~\text{$\phi $8}==\text{PhiBottom})\newline
~~~~~~~~~~~~)\newline
~~~~~~~~~~~~\&\&~~\text{NoPhiOnTop}~~\&\&~~\text{all$\phi $}==2*\text{PhiBottom}
\newline
~~~~~~~~~)
\newline\newline
~~~~~~~~~\|(\newline
~~~~~~~~~~~~(\newline
~~~~~~~~~~~~\text{(*The fully extended tesseract based adinkra}\newline
~~~~~~~~~~~~\text{(i.e. the real scalar supermultiplet) is included with this statement.*)}\newline
~~~~~~~~~~~~~~~(\text{$\phi $1$\psi $s}==4*\text{PsiDown}~~\&\&~~\text{$\phi $1}==\text{PhiBottom})\newline
~~~~~~~~~~~~~~~\&\&~~(\text{$\phi $2$\psi $s}==4*\text{PsiDown}~~\&\&~~\text{$\phi $2}==\text{PhiBottom})\newline
~~~~~~~~~~~~~~~\&\&~~\text{NoPhiOnTop}~~\&\&~~\text{all$\phi $}==2*\text{PhiBottom}))\newline
~~~~~~~)\newline
\newline
~~~~~~~\text{(*Exactly 2 fermion nodes are up}\newline
~~~~~~~\&\&~~\text{$\psi $all}\text{==}2*\text{PsiUp}
\newline
~~~)
\newline
~~~\|(\newline
~~~\text{(*2 boson nodes on the bottom row, and}\newline
~~~\text{ 1 or 2 fermion nodes raised and lowered*)}\newline
~~~~~~\text{$\phi $1}==\text{PhiBottom}~~\&\&~~\text{$\phi $1$\psi $s}==4*\text{PsiDown}\newline
~~~~~~\&\&~~\text{$\phi $8}==\text{PhiTop}~~\&\&~~\text{$\phi $8$\psi $s}==4*\text{PsiUp}\newline
~~~~~~\&\&~~(\text{Abs}[\text{$\phi $2}]+\text{Abs}[\text{$\phi $3}]+\text{Abs}[\text{$\phi $4}]\newline
~~~~~~+\text{Abs}[\text{$\phi $5}]+\text{Abs}[\text{$\phi $6}]+\text{Abs}[\text{$\phi $7}])==6*\text{PhiMid}\newline
~~~)
\newline
~~~\|(\newline
~~~~~~\text{$\phi $2}==\text{PhiBottom}~~\&\&~~\text{$\phi $2$\psi $s}==4*\text{PsiDown}\newline
~~~~~~\&\&~~\text{$\phi $7}==\text{PhiTop}~~\&\&~~\text{$\phi $7$\psi $s}==4*\text{PsiUp}\newline
~~~~~~\&\&~~(\text{Abs}[\text{$\phi $1}]+\text{Abs}[\text{$\phi $3}]+\text{Abs}[\text{$\phi $4}]\newline
~~~~~~+\text{Abs}[\text{$\phi $5}]+\text{Abs}[\text{$\phi $6}]+\text{Abs}[\text{$\phi $8}])==6*\text{PhiMid}\newline
~~~)
\newline
~~~\|(\newline
~~~~~~\text{$\phi $3}==\text{PhiBottom}~~\&\&~~\text{$\phi $3$\psi $s}==4*\text{PsiDown}\newline
~~~~~~\&\&~~\text{$\phi $6}==\text{PhiTop}~~\&\&~~\text{$\phi $6$\psi $s}==4*\text{PsiUp}\newline
~~~~~~\&\&~~(\text{Abs}[\text{$\phi $1}]+\text{Abs}[\text{$\phi $2}]+\text{Abs}[\text{$\phi $4}]\newline
~~~~~~+\text{Abs}[\text{$\phi $5}]+\text{Abs}[\text{$\phi $7}]+\text{Abs}[\text{$\phi $8}])==6*\text{PhiMid}\newline
~~~)
\newline
~~~\|(\newline
~~~~~~\text{$\phi $4}==\text{PhiBottom}~~\&\&~~\text{$\phi $4$\psi $s}==4*\text{PsiDown}\newline
~~~~~~\&\&~~\text{$\phi $5}==\text{PhiTop}~~\&\&~~\text{$\phi $5$\psi $s}==4*\text{PsiUp}\newline
~~~~~~\&\&~~(\text{Abs}[\text{$\phi $1}]+\text{Abs}[\text{$\phi $2}]+\text{Abs}[\text{$\phi $3}]\newline
~~~~~~+\text{Abs}[\text{$\phi $6}]+\text{Abs}[\text{$\phi $7}]+\text{Abs}[\text{$\phi $8}])==6*\text{PhiMid}\newline
~~~)\newline
$
\newline\newline
\text{(*Below we define the color matrices $\mathcal{B}_{iL}$ and $\mathcal{B}_{jR}$ for every adinkra*)}
\newline
greenL = ConstantArray[0, $\{8, 8\}$];\newline
orangeL = ConstantArray[0, $\{8, 8\}$];\newline
redL = ConstantArray[0, $\{8, 8\}$];\newline
purpleL = ConstantArray[0,$ \{8, 8\}$];\newline
\newline
greenR = ConstantArray[0, $\{8, 8\}$];\newline
orangeR = ConstantArray[0,$\{8, 8\}$];\newline
redR = ConstantArray[0, $\{8, 8\}$];\newline
purpleR = ConstantArray[0, $\{8, 8\}$];\newline

PosMat = ConstantArray[0, $\{5, 8\}$];

(*The below code makes the color matrices, and position matrix (PosMat).
 Here the position matrix shows the relative position of fermion and boson nodes by raising or lowering nodes from 
 the $\mathcal{N}=4$ valise adinkra in AdinkraMat v. 1.1. Thus a position matrix:\newline\newline
 $
 \left(
\begin{array}{cccccccc}
 0 & 0 & 0 & 0 & 0 & 0 & 0 & 0 \\
 0 & 0 & 0 & 0 & 0 & 0 & 0 & 0 \\
 0 & 0 & 0 & 0 & 1 & 1 & 1 & 1 \\
 \mathit{i} & \mathit{i} & \mathit{i} & \mathit{i} & \mathit{i} & \mathit{i} & \mathit{i} & \mathit{i} \\
 1 & 1 & 1 & 1 & 0 & 0 & 0 & 0
\end{array}
\right)
 $\newline\newline
 represents the 4 boson nodes, $\phi_5$, $\phi_6$, $\phi_7$, and  $\phi_8$ on the top row, with
 all the other boson nodes on the bottom row. The fermions are all in the center row.
  \newline
To make the code need fewer for loops, I added $\phi$ and/or $\psi$
together. Hence, if $\phi_8$ is above $\psi_8$, then the code knows that ``$\phi$8+$\psi$8=\text{PsiDown+PhiMid}=0", which implies the color matrices'
 elements for that node are $\b_i^{+1}$.

*)
\newline\newline
(*$\psi$8*)
\newline
purpleL[[8,  4]] =  $\b4\wedge((-1)\wedge(\psi 8 + \phi 4));$(*connects to $\phi$4*)\newline
redL[[8, 6]] =$  \b3\wedge((-1)\wedge(\psi8 + \phi6));$(*connects to
$\phi$6*)\newline
orangeL[[8,   7]] =$ \b2\wedge((-1)\wedge(\psi8 + \phi7));$(*connects to $\phi$7*)\newline
greenL[[8, 
  8]] =$ \b1\wedge((-1)\wedge(\phi8 + \psi8));$(*connects to $\phi$8*)\newline
PosMat[[4 - 2*$\psi$8, 8]] =$ \dot{\imath};$
\newline
\newline
(*$\psi$7*)
\newline
purpleL[[7, 
  3]] =$  \b4\wedge((-1)\wedge(\phi3 + \psi7));$(*connects to $\phi$3*)\newline
redL[[7, 5]] =$ \b3\wedge((-1)\wedge(\phi5 + \psi7));$(*connects to \
$\phi$5*)\newline
greenL[[7, 
  7]] =$ \b1\wedge((-1)\wedge(\phi7 + \psi7));$(*connects to $\phi$7*)\newline
orangeL[[7, 
  8]] =$ \b2\wedge((-1)\wedge(\phi8 + \psi7));$(*connects to $\phi$8*)\newline
PosMat[[4 - 2*$\psi$7, 7]] =$ \dot{\imath};$
\newline
\newline
(*$\psi$6*)
\newline
purpleL[[6, 
  2]] =$ \b4\wedge((-1)\wedge(\phi2 + \psi6));$(*connects to $\phi$2*)\newline
orangeL[[6, 
  5]] =$ \b2\wedge((-1)\wedge(\phi5 + \psi6));$(*connects to $\phi$5*)\newline
greenL[[6, 
  6]] =$ \b1\wedge((-1)\wedge(\phi6 + \psi6));$(*connects to $\phi$6*)\newline
redL[[6, 8]] =$ \b3\wedge((-1)\wedge(\phi8 + \psi6));$(*connects to \
$\phi$8*)\newline
PosMat[[4 - 2*$\psi$6, 6]] =$ \dot{\imath};$
\newline
\newline
(*$\psi$5*)
\newline
purpleL[[5, 
  1]] =$ \b4\wedge((-1)\wedge(\phi1 + \psi5));$(*connects to $\phi$1*)\newline
greenL[[5, 
  5]] =$ \b1\wedge((-1)\wedge(\phi5 + \psi5));$(*connects to $\phi$5*)\newline
orangeL[[5, 
  6]] =$ \b2\wedge((-1)\wedge(\phi6 + \psi5));$(*connects to $\phi$6*)\newline
redL[[5, 7]] =$ \b3\wedge((-1)\wedge(\phi7 + \psi5));$(*connects to \
$\phi$7*)\newline
PosMat[[4 - 2*$\psi$5, 5]] =$ \dot{\imath};$
\newline
\newline
(*$\psi$4*)
\newline
redL[[4, 2]] =$ \b3\wedge((-1)\wedge(\psi4 + \phi2));$(*connects to \
$\phi$2*)\newline
orangeL[[4, 
  3]] =$ \b2\wedge((-1)\wedge(\psi4 + \phi3));$(*connects to $\phi$3*)\newline
greenL[[4, 
  4]] =$ \b1\wedge((-1)\wedge(\psi4 + \phi4));$(*connects to $\phi$4*)\newline
purpleL[[4, 
  8]] =$ \b4\wedge((-1)\wedge(\phi8 + \psi4));$(*connects to $\phi$8*)\newline
PosMat[[4 - 2*$\psi$4, 4]] =$ \dot{\imath};$
\newline
\newline
(*$\psi$3*)
\newline
redL[[3, 1]] =$ \b3\wedge((-1)\wedge(\phi1 + \psi3));$(*connects to \
$\phi$1*)\newline
greenL[[3, 
  3]] =$ \b1\wedge((-1)\wedge(\psi3 + \phi3));$(*connects to $\phi$3*)\newline
orangeL[[3, 
  4]] =$ \b2\wedge((-1)\wedge(\psi3 + \phi4));$(*connects to $\phi$4*)\newline
purpleL[[3, 
  7]] =$ \b4\wedge((-1)\wedge(\psi3 + \phi7));$(*connects to $\phi$7*)\newline
PosMat[[4 - 2*$\psi$3, 3]] =$ \dot{\imath};$
\newline
\newline
(*$\psi$2*)
\newline
orangeL[[2, 
  1]] =$ \b2\wedge((-1)\wedge(\psi2 + \phi1));$(*connects to $\phi$1*)\newline
greenL[[2, 
  2]] =$ \b1\wedge((-1)\wedge(\psi2 + \phi2));$(*connects to $\phi$2*)\newline
redL[[2, 4]] =$ \b3\wedge((-1)\wedge(\psi2 + \phi4));$(*connects to \
$\phi$4*)\newline
purpleL[[2, 
  6]] =$ \b4\wedge((-1)\wedge(\psi2 + \phi6));$(*connects to $\phi$6*)\newline
PosMat[[4 - 2*$\psi$2, 2]] =$ \dot{\imath};$
\newline
\newline
(*$\psi$1*)
\newline
greenL[[1, 
  1]] =$ \b1\wedge((-1)\wedge(\phi1 + \psi1));$(*connects to $\phi$1*)\newline
orangeL[[1, 
  2]] =$ \b2\wedge((-1)\wedge(\psi1 + \phi2));$(*connects to $\phi$2*)\newline
redL[[1, 3]] =$ \b3\wedge((-1)\wedge(\psi1 + \phi3));$(*connects to \
$\phi$3*)\newline
purpleL[[1, 
  5]] =$ \b4\wedge((-1)\wedge(\psi1 + \phi5));$(*connects to $\phi$5*)\newline
PosMat[[4 - 2*$\psi$1, 1]] =$ \dot{\imath};$
\newline
\newline
(*$\phi$8*)
\newline
greenR[[8, 
  8]] =$ \b1\wedge((-1)\wedge(1 + \phi8 + \psi8));$(*connects to $\psi$8*)\newline
orangeR[[8, 
  7]] =$ \b2\wedge((-1)\wedge(1 + \phi8 + \psi7));$(*connects to $\psi$7*)\newline
redR[[8, 6]] =$  \b3\wedge((-1)\wedge(1 + \phi8 + \psi6));$(*connects to \
$\psi$6*)\newline
purpleR[[8, 
  4]] =$  \b4\wedge((-1)\wedge(1 + \phi8 + \psi4));$(*connects to \
$\psi$4*)\newline
PosMat[[3 - 2*$\phi$8, 8]] =$ 1;$
\newline
\newline
(*$\phi$7*)
\newline
greenR[[7, 
  7]] =$ \b1\wedge((-1)\wedge(1 + \phi7 + \psi7));$(*connects to $\psi$7*)\newline
orangeR[[7, 
  8]] =$ \b2\wedge((-1)\wedge(\psi8 + 1 + \phi7));$(*connects to $\psi$8*)\newline
purpleR[[7, 
  3]] =$  \b4\wedge((-1)\wedge(\psi3 + 
      1 + \phi7));$(*connects to $\psi$3*)\newline
redR[[7, 5]] =$ \b3\wedge((-1)\wedge(\psi5 + 
      1 + \phi7));$(*connects to $\psi$5*)\newline
PosMat[[3 - 2*$\phi$7, 7]] =$ 1;$
\newline
\newline
(*$\phi$6*)
\newline
greenR[[6, 
  6]] =$ \b1\wedge((-1)\wedge(1 + \phi6 + \psi6));$(*connects to $\psi$6*)\newline
redR[[6, 8]] =$ \b3\wedge((-1)\wedge(\psi8 + 
      1 + \phi6));$(*connects to $\psi$8*)\newline
purpleR[[6, 
  2]] =$ \b4\wedge((-1)\wedge(\psi2 + 1 + \phi6));$(*connects to $\psi$2*)\newline
orangeR[[6, 
  5]] =$ \b2\wedge((-1)\wedge(\psi5 + 1 + \phi6));$(*connects to $\psi$5*)\newline
PosMat[[3 - 2*$\phi$6, 6]] =$ 1;$
\newline
\newline
(*$\phi$5*)
\newline
orangeR[[5, 
  6]] =$ \b2\wedge((-1)\wedge(\psi6 + 1 + \phi5));$(*connects to $\psi$6*)\newline
redR[[5, 7]] =$ \b3\wedge((-1)\wedge(\psi7 + 
      1 + \phi5));$(*connects to $\psi$7*)\newline
purpleR[[5, 
  1]] =$ \b4\wedge((-1)\wedge(\psi1 + 1 + \phi5));$(*connects to $\psi$1*)\newline
greenR[[5, 
  5]] =$ \b1\wedge((-1)\wedge(1 + \phi5 + \psi5));$(*connects to $\psi$5*)\newline
PosMat[[3 - 2*$\phi$5, 5]] =$ 1;$
\newline
\newline
(*$\phi$4*)
\newline
orangeR[[4, 
  3]] =$ \b2\wedge((-1)\wedge(1 + \phi4 + \psi3));$(*connects to $\psi$3*)\newline
redR[[4, 2]] =$ \b3\wedge((-1)\wedge(1 + \phi4 + \psi2));$(*connects to \
$\psi$2*)\newline
purpleR[[4, 
  8]] =$ \b4\wedge((-1)\wedge(\psi8 + 1 + \phi4));$(*connects to $\psi$8*)\newline
greenR[[4, 
  4]] =$ \b1\wedge((-1)\wedge(1 + \phi4 + \psi4));$(*connects to $\psi$4*)\newline
PosMat[[3 - 2*$\phi$4, 4]] =$ 1;$
\newline
\newline
(*$\phi$3*)
\newline
greenR[[3, 
  3]] =$ \b1\wedge((-1)\wedge(1 + \phi3 + \psi3));$(*connects to $\psi$3*)\newline
redR[[3, 1]] =$ \b3\wedge((-1)\wedge(\psi1 + 
      1 + \phi3));$(*connects to $\psi$1*)\newline
purpleR[[3, 
  7]] =$ \b4\wedge((-1)\wedge(1 + \phi3 + \psi7));$(*connects to $\psi$7*)\newline
orangeR[[3, 
  4]] =$ \b2\wedge((-1)\wedge(1 + \phi3 + \psi4));$(*connects to $\psi$4*)\newline
PosMat[[3 - 2*$\phi$3, 3]] =$ 1;$
\newline
\newline
(*$\phi$2*)
\newline
greenR[[2, 
  2]] =$ \b1\wedge((-1)\wedge(1 + \phi2 + \psi2));$(*connects to $\psi$2*)\newline
orangeR[[2, 
  1]] =$ \b2\wedge((-1)\wedge(1 + \phi2 + \psi1));$(*connects to $\psi$1*)\newline
purpleR[[2, 
  6]] =$ \b4\wedge((-1)\wedge(1 + \phi2 + \psi6));$(*connects to $\psi$6*)\newline
redR[[2, 4]] =$ \b3\wedge((-1)\wedge(1 + \phi2 + \psi4));$(*connects to \
$\psi$4*)\newline
PosMat[[3 - 2*$\phi$2, 2]] =$ 1;$
\newline
\newline
(*$\phi$1*)
\newline
greenR[[1, 
  1]] =$ \b1\wedge((-1)\wedge(1 + \phi1 + \psi1));$(*connects to $\psi$1*)\newline
orangeR[[1, 
  2]] =$ \b2\wedge((-1)\wedge(1 + \phi1 + \psi2));$(*connects to $\psi$2*)\newline
redR[[1, 3]] =$ \b3\wedge((-1)\wedge(1 + \phi1 + \psi3));$(*connects to \
$\psi$3*)\newline
purpleR[[1, 
  5]] =$ \b4\wedge((-1)\wedge(1 + \phi1 + \psi5));$(*connects to $\psi$5*)\newline
PosMat[[3 - 2*$\phi$1, 1]] = 1;
\newline\newline
AdinkraNum++;(*Counts the number of adinkras we have looked through*)\newline
\newline
(*We organizing all the matrices into an array, GORP, which is 
an acronym for the line colors. Once all liftable adinkras are found, 
one may glean all necessary information of an adinkra
by plugging in its associated number into the
``AdinkraNum" index of GORP.*)
\newline
\newline
GORP[[1, All, All, AdinkraNum]] = PosMat;\newline
GORP[[2, All, All, AdinkraNum]] = redL.orangeR;\newline
GORP[[3, All, All, AdinkraNum]] = redL.greenR;\newline
GORP[[4, All, All, AdinkraNum]] = redL.purpleR;\newline
GORP[[5, All, All, AdinkraNum]] = orangeL.greenR;\newline
GORP[[6, All, All, AdinkraNum]] = orangeL.purpleR;\newline
GORP[[7, All, All, AdinkraNum]] = greenL.purpleR;\newline

(*Eigenvalues of color matrices used to find bow ties*)
\newline\newline
Evals[[RedOrange, AdinkraNum]] = Eigenvalues[redL.orangeR];\newline
Evals[[RedGreen, AdinkraNum]] = Eigenvalues[redL.greenR];\newline
Evals[[RedPurple, AdinkraNum]] = Eigenvalues[redL.purpleR];\newline
Evals[[OrangeGreen, AdinkraNum]] = Eigenvalues[orangeL.greenR];\newline
Evals[[OrangePurple, AdinkraNum]] = Eigenvalues[orangeL.purpleR];\newline
Evals[[GreenPurple, AdinkraNum]] = Eigenvalues[ greenL.purpleR];\newline
(*

When determining liftability, we remember that,
If color i and j form bow ties, then we put the colors in a set S.
for any other color - if it forms bow ties with colors in S, then it
is also in S.\newline

If there exists a color (or colors) not in S (i.e. in another set S') \
then the adinkra is liftable\newline
*)\newline\newline
BowTies = ConstantArray[False, 6];\newline
(*Here we check if the given color combo forms any bow ties. *)\newline
For[ColorCombo = 1, ColorCombo $\leq$ 6, ColorCombo++,\newline
$~~~$For[evalIndex = 1, evalIndex $\leq$ Dimensions[Evals[[ColorCombo, AdinkraNum]]][[1]], 
$~~~~~~~$ evalIndex++,\newline
$~~~~~~$(*Two color bow tie formed. put both colors in the set BowTies:\newline
$~~~~~~~~~$1: red-orange\newline
$~~~~~~~~~$2: red-green\newline
$~~~~~~~~~$3: red-purple\newline
$~~~~~~~~~$4: orange-green\newline
$~~~~~~~~~$5:orange-purple\newline
$~~~~~~~~~$6:green-purple\newline
 $~~~~~~$*)\newline
$~~~~~~$If[Abs[Evals[[ColorCombo, AdinkraNum]][[evalIndex]]] $\neq$ 1,\newline 
$~~~~~~$BowTies[[ColorCombo]] = True;\newline
$~~~~~~$Break[];\newline
$~~~~~~$]
]
 ]
\newline\newline
(*Cases when the adinkras are liftable:\newline\newline
 $~~~$1: Green colored lines are bow tie free\newline
 $~~~$2: Orange colored lines are bow tie free\newline
 $~~~$3: Red colored lines are bow tie free\newline
 $~~~$4:Purple colored lines are bow tie free\newline
 $~~~$5: \emph{Only} red-orange and green-purple bow ties\newline
 $~~~$6:  \emph{Only} red-purple and green-orange bow ties\newline
 $~~~$7:  \emph{Only} red-green and orange-purple bow ties\newline
*)
\newline
\newline
If[(*case 1*)\newline
  BowTies[[GreenRed]] == False \newline
  $\&\&$ BowTies[[GreenOrange]] == False\newline 
  $\&\&$ BowTies[[GreenPurple]] == False,\newline
 $~~~$numLiftable++;(*Here we count the number of \newline
 $~~~~~~~~~~~~~~~~~~~~~~~~$(un-unique) liftable adinkras found so far*)\newline
 $~~~$LiftableAdinkraNum[[numLiftable]] = AdinkraNum;\newline
 $~~~~~~~~~~~~~~~~~~~~~~$(*Here we put the liftable \newline
 $~~~~~~~~~~~~~~~~~~~~~~$adinkra's label into a convenient array*)\newline
 Break[];\newline
 ]
 \newline
 \newline
If[(*case 2*)\newline
 BowTies[[OrangeRed]] == False \newline
 $\&\&$ BowTies[[OrangeGreen]] == False\newline 
 $\&\&$  BowTies[[OrangePurple]] == False,\newline
 $~~~$numLiftable++;\newline
 $~~~$LiftableAdinkraNum[[numLiftable]] = AdinkraNum;\newline
 $~~~$Break[];\newline
 ]
 \newline
 \newline
If[(*case 3*)\newline
 BowTies[[RedGreen]] == False \newline
 $\&\&$ BowTies[[RedOrange]] == False \newline
 $\&\&$  BowTies[[RedPurple]] == False,\newline
 $~~~$numLiftable++;\newline
 $~~~$LiftableAdinkraNum[[numLiftable]] = AdinkraNum;\newline
 $~~~$Break[];\newline
 ]
\newline
\newline
If[(*case 4*)\newline
BowTies[[PurpleRed]] == False \newline
$\&\&$ BowTies[[PurpleOrange]] == False \newline
$\&\&$ BowTies[[PurpleGreen]] == False,\newline
$~~~$ numLiftable++;\newline
$~~~$ LiftableAdinkraNum[[numLiftable]] = AdinkraNum;\newline
$~~~$ Break[];\newline
 ]
\newline
\newline
If[(*case 5*)\newline
 BowTies[[RedOrange]] == True \newline
 $\&\&$ BowTies[[GreenPurple]] == True \newline
 $\&\&$  Sum[BowTies[[ii]], $\{$ii, 1, 6$\}$] == 2*True + 4*False,\newline
 $~~~$numLiftable++;\newline
 $~~~$LiftableAdinkraNum[[numLiftable]] = AdinkraNum;\newline
 $~~~$Break[];\newline
 ]
\newline
\newline
If[(*case 6*)\newline 
 BowTies[[RedPurple]] == True \newline
 $\&\&$ BowTies[[OrangeGreen]] == True\newline 
 $\&\&$  Sum[BowTies[[ii]], $\{$ii, 1, 6$\}$] == 2*True + 4*False,\newline
 $~~~$numLiftable++;\newline
 $~~~$LiftableAdinkraNum[[numLiftable]] = AdinkraNum;\newline
 $~~~$Break[];\newline
 ]
 \newline
 \newline
If[(*case 7*)\newline 
 BowTies[[RedGreen]] == True\newline
  $\&\&$ BowTies[[OrangePurple]] == True\newline 
  $\&\&$ Sum[BowTies[[ii]], $\{$ii, 1, 6$\}$] == 2*True + 4*False,\newline
 $~~~$numLiftable++;\newline
 $~~~$LiftableAdinkraNum[[numLiftable]] = AdinkraNum;\newline
 $~~~$Break[];\newline
]
\newline\newline
]]]]]]]]]]]]]]]]](*These are end-brackets for the for loops*)\newline\newline
(*

Lastly, we look at all the adinkras found, and determine which are unique. In the $\mathcal{N}=4$ case, this entails looking at whether we have the same number of $(I,J)$ bow ties for any give colors $I$ and $J$. $(3|7|5|1)$ adinkras are not counted with our method, because they are just the flipped version of $(1|5|7|3)$ adinkras., hence knowing one uniquely determines the other.
\newline

The algorithm is as follows:\newline\newline
$~~~$-Look at the total liftable adinkras.\newline
$~~~$-First liftable adinkra is the first ``unique" one\newline
$~~~$-This is in the list of ``unique adinkras so far found"\newline
$~~~$-For every new adinkra, compare (sorted) bow tie eigenvalues\newline
$~~~$ with the previous unique adinkras\newline
$~~~$-If the eigenvalues are the same, ignore (we found \newline
$~~~$that knowing the total number of bowties can go a long way \newline
$~~~$to determining whether two adinkras are the same).\newline\newline
*)\newline\newline
TotalUniqueAdinkras = 1;\newline
(*LiftableAdinkraNum[[1]]*)\newline
UniqueAdinkraNum = ConstantArray[0, 18];\newline
UniqueAdinkraNum[[1]] = LiftableAdinkraNum[[1]];\newline
\newline
For[i = 2, i $\le$ numLiftable, i++,\newline
$~~~$count = 0;(*We use this variable to determine whether an adinkra is unique*)\newline
$~~~$For[j = 1, j $\le$ TotalUniqueAdinkras, j++,\newline 
$~~~~~~$NewAdnk = \newline
$~~~~~~~~~$Sort[Join[Evals[[1, LiftableAdinkraNum[[i]]]], \newline
$~~~~~~~~~$Evals[[2, LiftableAdinkraNum[[i]]]], \newline
$~~~~~~~~~$Evals[[3, LiftableAdinkraNum[[i]]]], \newline
$~~~~~~~~~$Evals[[4, LiftableAdinkraNum[[i]]]], \newline
$~~~~~~~~~$Evals[[5, LiftableAdinkraNum[[i]]]],\newline
$~~~~~~~~~$ Evals[[6, LiftableAdinkraNum[[i]]]]]];\newline
$~~~~~~$OldAdnk = \newline
$~~~~~~~~~$Sort[Join[Evals[[1, UniqueAdinkraNum[[j]]]], \newline
$~~~~~~~~~$Evals[[2, UniqueAdinkraNum[[j]]]], \newline
$~~~~~~~~~$Evals[[3, UniqueAdinkraNum[[j]]]], \newline
$~~~~~~~~~$Evals[[4, UniqueAdinkraNum[[j]]]], \newline
$~~~~~~~~~$Evals[[5, UniqueAdinkraNum[[j]]]],\newline
$~~~~~~~~~$Evals[[6, UniqueAdinkraNum[[j]]]]]];\newline
$~~~~~~$If[ NewAdnk == OldAdnk,\newline
$~~~~~~~~~$Break[];,\newline
$~~~~~~~~~$(*If not the same as the previous adinkra*)\newline
$~~~~~~~~~$count++;\newline
$~~~~~~$]\newline     
 $~~~$]
 \newline
  (*If unique*)\newline
$~~~$If[count == TotalUniqueAdinkras,\newline
$~~~~~~$TotalUniqueAdinkras++;\newline
$~~~~~~$ UniqueAdinkraNum[[TotalUniqueAdinkras]] = LiftableAdinkraNum[[i]];\newline
 $~~~$]\newline
 ]
$$~~$$

\end{document}